\documentclass[agums]{aguplus}
\usepackage{epsfig}

\everymath={\fam0 }

\righthead{Marine Boundary Layer Ozone} 
\lefthead{\it Lary}

\authoraddr{K.-Y. Wang, Centre for Atmospheric Science, Department of
Chemistry, Cambridge University, Lensfield Road, Cambridge, CB2 1EW,
U.K.}

\authoraddr{D.E. Shallcross, School of Chemistry, University of Bristol,
BS8 1TS, U.K.}

\authoraddr{J.A. Pyle, Centre for Atmospheric Science, Department of
Chemistry, Cambridge University, Lensfield Road, Cambridge, CB2 1EW,
U.K.}

\begin{document}

\title{Modelling the spring ozone maximum and the interhemispheric asymmetry 
in the remote marine boundary layer \\
1. Comparison with surface and ozonesonde measurements}

\author{\altaffilmark{1}K.-Y. Wang, \altaffilmark{2}D.E. Shallcross, and  \altaffilmark{3}J.A. Pyle}
\affil{1. Department of Atmospheric Sciences, National Central University, Chung-Li, Taiwan}
\affil{2. Centre for Biogeochemistry, School of Chemistry, Bristol University, BS8 1TS, U.K.}
\affil{3. Centre for Atmospheric Science, Cambridge University, CB2 1EW, U.K.}

\begin{abstract}
Here we report a modelling study of the
spring ozone maximum and its
interhemispheric asymmetry in the remote marine boundary layer (MBL).
The modelled results
are examined at the surface and on a series of time-height cross sections
at several locations spread over the Atlantic, the Indian, and the Pacific Oceans.
Comparison of model with surface measurements at remote MBL stations
indicate a close agreement. 
The most striking feature of the hemispheric
spring ozone maximum in the MBL can be most easily identified
at the NH sites of Westman Island,
Bermuda, and Mauna Loa, and at the SH site of Samoa. 
Modelled ozone vertical distributions in the troposphere are compared with 
ozone profiles. 
For the Atlantic and the Indian sites, 
the model generally produces a hemispheric spring
ozone maximum close to those of the measurements.
The model also produces a spring ozone maximum
in the northeastern and tropical north Pacific close to those measurements,
and at sites in the NH high latitudes. 
The good agreement between model and measurements indicate that the model
can reproduce the proposed mechanisms responsible for producing the
spring ozone maximum in these regions of the MBL, 
lending confidence in the use of the model to investigate MBL ozone chemistry
(see part 2 and part 3). The
spring ozone maximum in the tropical central south Pacific and eastern
equatorial Pacific are less well reproduced by the model, indicating that
both the transport of $O_3$ precursors from 
biomass burning emissions taking place in southeastern Asia, Australia, Oceania,
southern Africa, and South America are not well represented in the model
in these regions. 
Overall, the model produces a better simulation at sites where the stratosphere
and biomass burning emissions are the major contributors. 
\vskip  0.1in
{\it Key word index:} ozone, CTM, spring ozone maximum, interhemispheric asymmetry, biomass burning

\end{abstract}

%
% 1. Introduction
% 
\section{1. Introduction}

%
% Why do we want to do this study?
%

% Why do we need to care about ozone?

Due to its impact on human health [e.g., {\it K\"{u}nzli et al.}, 2000]
and terrestrial vegetation [e.g., {\it Finnan et al.}, 1996], and its role
as a greenhouse gas,
ozone has long 
been intensively studied via surface and ozonesonde measurements, 
aircraft observations, satellite total ozone mapping, and numerical modelling.
%
% Surface measurements 
%
Many long-term and intensive surface ozone measurements in the remote marine environment  
had been conducted over
the Atlantic
[e.g., {\it Winkler}, 1988; {\it Penkett et al.}, 1998; {\it Junkermann and Stockwell}, 1999], 
the Pacific
[e.g., {\it Piotrowicz et al.}, 1986; {\it Piotrowicz et al.}, 1991; 
{\it Jaffe et al.}, 1996; {\it Crawford et al.}, 1997;  {\it Kajii et al.}, 1997;
{\it Pochanart et al.}, 1999; {\it Monks et al.}, 2000],
and the Indian Ocean 
[e.g., {\it Lal et al.}, 1998].
%
% Ozonesonde measurements
%  - must quote Logan's paper
%
Since there are limitations in observations of ozone and other constituents primarily at the 
surface, many measurements such as 
ozonesonde 
[e.g., {\it Moody et al.} [1995]; {\it Oltmans et al.}, 1996; 
{\it Logan}, 1999, and references therein; {\it Latt et al.}, 1999],
and airplanes
[e.g., {\it Kawa and Pearson}, 1989; {\it Murphy and Fahey}, 1994]
were used to provide the necessary profile
information for understanding the chemistry and dynamics controlling the 
variations of ozone in the troposphere.

% What have been shown from these measurements?

{\it Monks} [2000] gave a comprehensive 
review of the surface observations and the springtime ozone maximum, while
{\it Logan} [1999] documented ozonesonde measurements and derived a
climatology for tropospheric ozone based on these sonde data.
Two prominent features constantly show up from these measurements.
First, the appearance of a spring maximum in the troposphere;
and secondly, the existence of an interhemispheric asymmetry between the
northern and southern hemispheres
[e.g., {\it Winkler}, 1988; {\it Johnson et al.}, 1990]. 
These interhemispheric asymmetries also featured in three-dimensional (3D) modelling
studies
[e.g., {\it M\"{u}ller and Brasseur}, 1995].

%
% Two competing theories
%
Basically there are two theories regarding the origins of elevated ozone 
in the remote troposphere. 
First, transport from the stratosphere to
the troposphere is the dominating source of ozone in the troposphere
[e.g., {\it Moody et al.}, 1995; {\it Oltmans et al.}, 1996; {\it Roelofs and Lelieveld}, 1997].
Second, that tropospheric ozone originated mainly from 
emissions (e.g., biomass burning emissions of $O_3$ precursors, and NOx
sources from soil and lightning in the continents), photochemistry
(photochemical oxidation of CO and hydrocarbons catalysed by HOx and NOx
[e.g., see {\it Monks}, 2000, and references therein]), and transport
processes (e.g., large-scale long-range transport, cloud convection,
and vertical mixing between the atmospheric boundary layer and free
troposphere) within the troposphere [{\it Roelofs et al.}, 1997; {\it Yinger et al.}, 1999; {\it Logan}, 1999]. 
%
% General consensus: non-indigenous and must be transported from elsewhere
%
The general consensus is that the elevated ozone concentrations 
in the remote MBL are non-indigenous, and the transport of either or
both elevated ozone and its precursors from elsewhere is the major contributing
factor for the observed high ozone levels in the remote MBL. 
%
%  Photochemical loss 
%
In addition,
very low ozone concentrations have also been observed in the tropical MBL 
[{\it Kley et al.}, 1996; {\it Singh et al.}, 1996], indicating that halogens may
play an important role in oxidation processes and the ozone budget in parts of 
the remote MBL 
[{\it Ariya et al.}, 1998; {\it Dickerson et al.}, 1999; {\it Nagao et al.}, 1999].

%
% What do we want to do in this study?
%

Hence modelling the spring ozone maximum and interhemispheric asymmetry
remains one of the most critical tests to our current 
understanding of tropospheric chemistry in the remote MBL
[{\it Winkler}, 1988; {\it Logan}, 1999; {\it Monks}, 2000].
The recently available ozonesonde measurements provide a much higher 
global coverage of the ozone vertical profiles 
from surface to the lower stratosphere; 
however, these ozonesondes have
not yet been widely used to address the issue such as spring ozone maximum
and interhemispheric asymmetry [e.g., {\it Logan}, 1999].
In this first part of a three-part series of papers, concerning the sources of
the spring ozone maximum and its interhemispheric asymmetry in the remote MBL,
we present results from a
3D chemistry transport model (CTM) and a detailed annual comparison with surface 
and ozonesonde measurements at several locations in the remote marine environments.
A modelling test on the {\it stratosphere-dominated theory} 
is discussed in the
second part of the paper. Finally, the test on the {\it self-contained theory}   
is discussed in the final part of the paper.

% How do I do this study (method)?
%
% 2. The IMS Model
%  
\section{2. The IMS Model}

A detailed description about the formulation and evaluation of model
emission inventories, transport processes, chemistry, and
the simulated $O_3$, $CH_4$, and CO distributions in the troposphere
were described in {\it Wang et al.} [2001], {\it Wang et al.} [1999],
and {\it Wang and Shallcross} [2000].
Briefly, the model
uses a semi-Lagrangian approach for the large-scale advection of
long-lived species. Vertical mixing
of species in the atmospheric boundary layer is modelled following the
radiatively driven diurnal
variation of the boundary-layer height.
Vertical redistribution of chemical species through cloud convection
is achieved using a mass-flux cloud scheme.
The model uses a comprehensive gas-phase reaction mechanism
for NOx, methane, NMHC, biogenic VOCs, and other sulfur and halogen
compounds. Specified emissions for anthropogenic sources is that of
EDGAR and GEIA. Geographical distributions of the
sinks of important tropospheric species ($O_3$, $NO_2$,
PAN, CO, etc) are considered via dry and wet deposition processes.
The model uses analyzed winds (1992) from ECMWF, and it contains 19 vertical layers
which extends from surface to about 10 hPa. The model horizontal resolution
is about $7.5^{\circ}$ in longitude and $4.5^{\circ}$ in latitude.

% What are the findings/new discoveries from this study (results)?
%
% 3. Results
%
\section{3. Results}

The ozone simulations were performed using the IMS model with
analyzed data of zonal wind, meridional wind, temperature,
specific humidity, and surface pressure from the European Centre
for Medium Range Weather Forecasts (ECMWF) which are updated every 6 hours.
The IMS model was multitasked
and run parallelly on the shared-memory CRAY J90 [{\it Wang et al.}, 2000].
The model was run for two years with full tropospheric chemistry,
and the results, composed from every 6-hour model output frequency,  
from the second year were used for the following discussion.
We note that the meteorology used here is not directly corresponding to
the years where the ozonesonde data were available. Hence, some 
variability in the distribution of tropospheric ozone can be caused by   
the interannual
variability in the tropics (eg., {\it Peters et al.}, 2001).

\subsection{3.1. Surface Ozone Distribution}

\begin{figure*}[hp]
\vbox{
\vskip -0.0in
\centerline{
\leavevmode
(a)
\epsfxsize=2.5in
\epsfysize=3.0in
\rotatebox{-90.}{\epsfbox{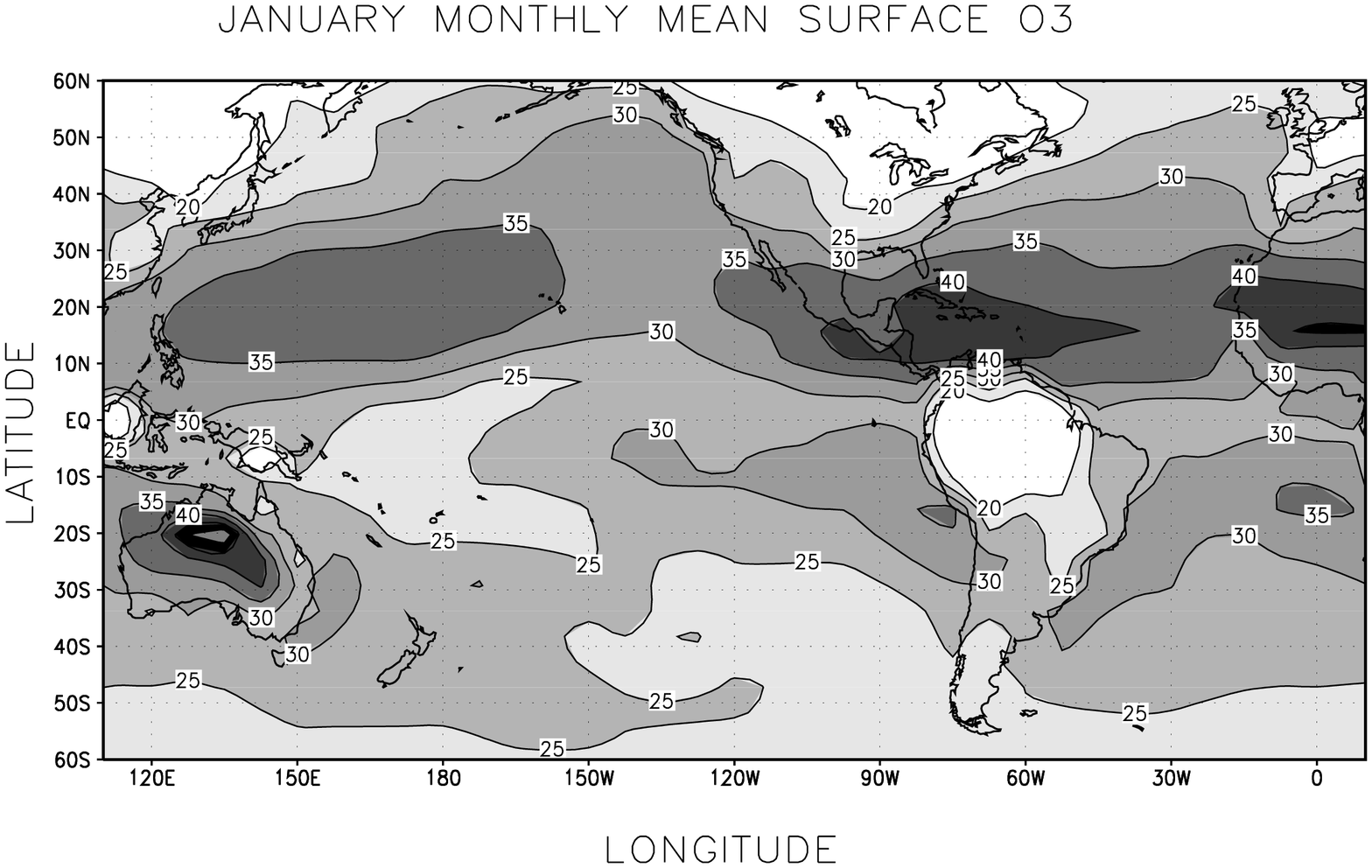}}
(b)
\epsfxsize=2.5in
\epsfysize=3.0in
\rotatebox{-90.}{\epsfbox{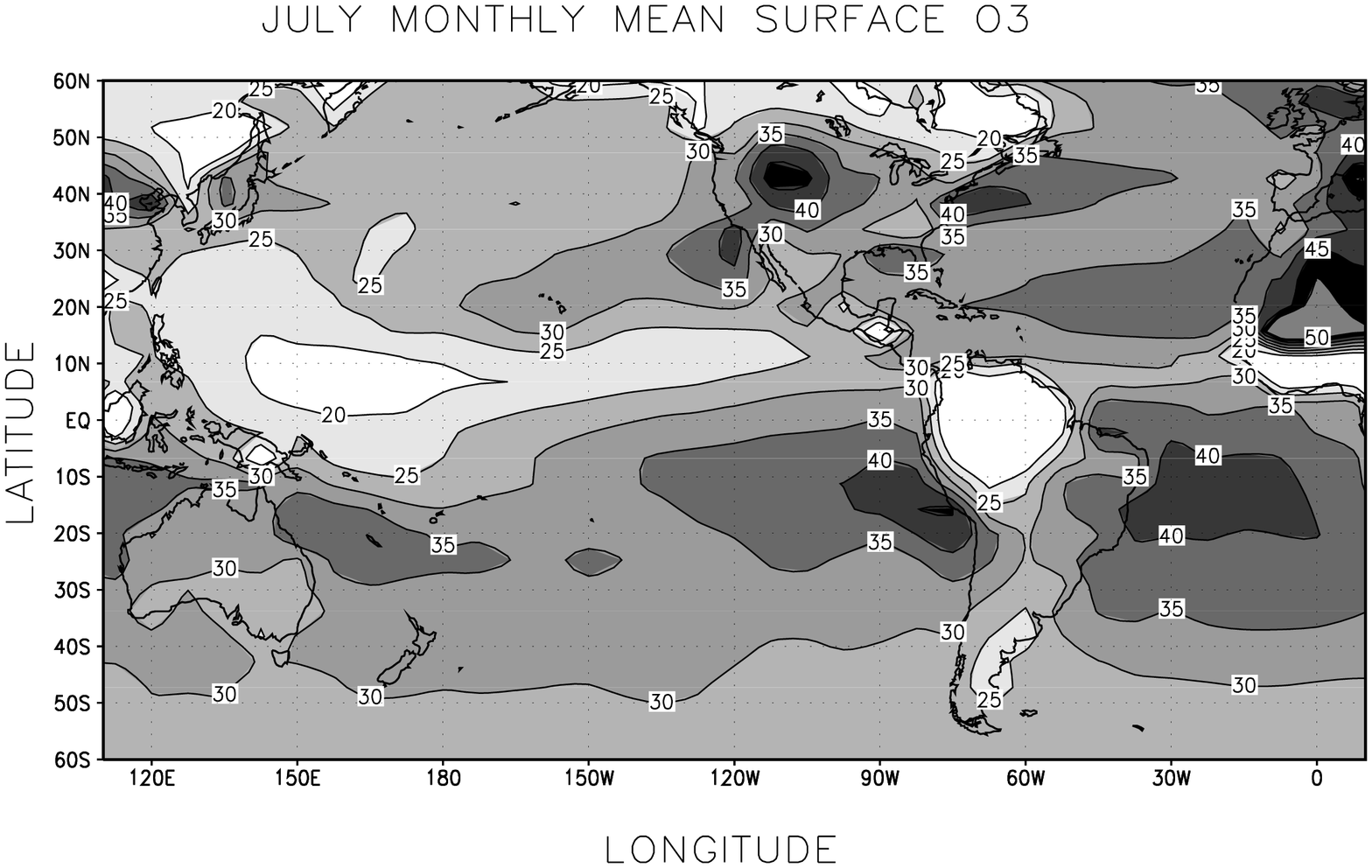}}
}
\vskip -0.0in
\centerline{
\leavevmode
(c)
\epsfxsize=3.0in
\epsfysize=6.0in
\rotatebox{-90.}{\epsfbox{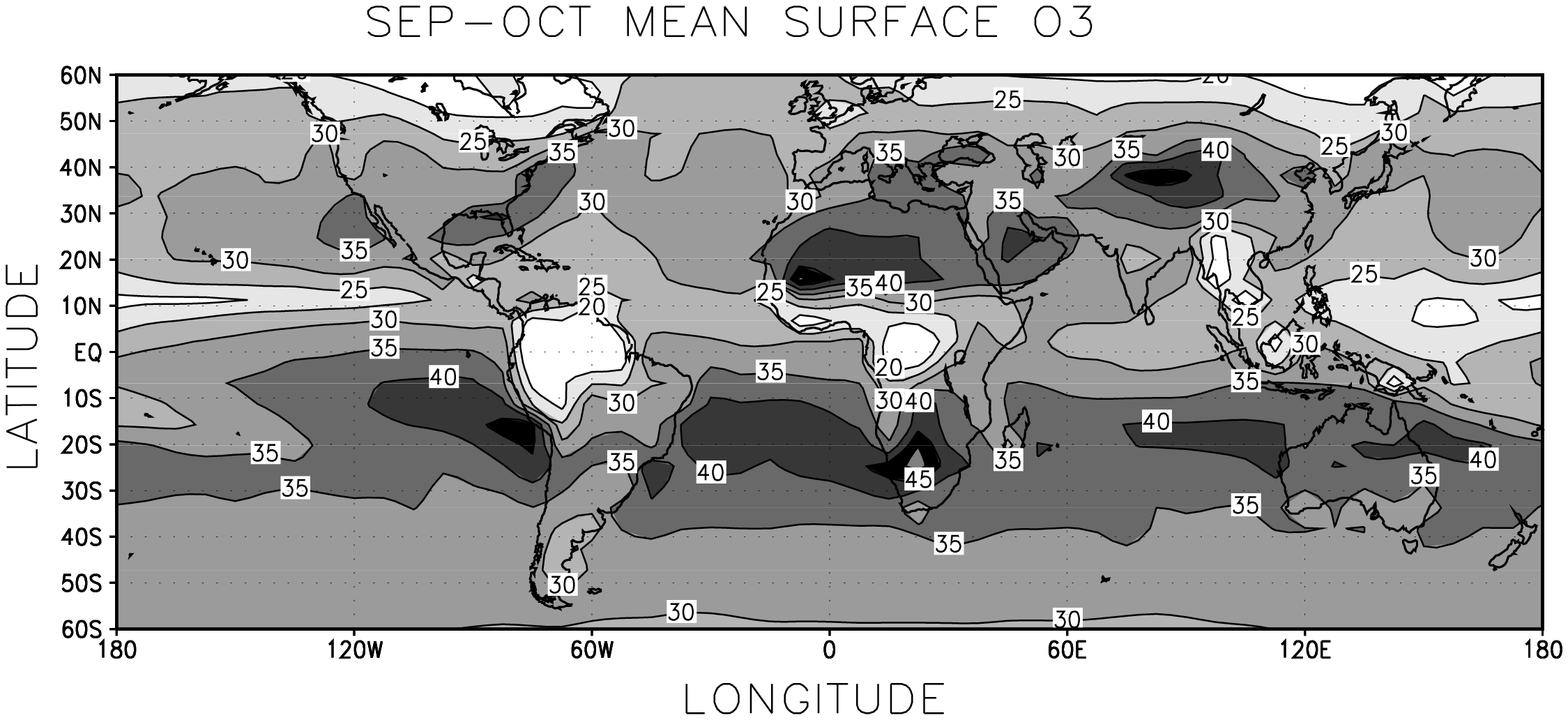}}
}
}
\caption{ \label{fig.8.1} Monthly mean ozone distributions (ppbv) calculated
at the surface for (a) January, (b) July, and (c) September to October.}
\end{figure*}

Figure~\ref{fig.8.1} shows modelled surface ozone distributions over most of
the marine boundary layer in the Atlantic and the Pacific 
(from $120^{\circ}E$ to $0^{\circ}W$, and from $60^{\circ}N$ to $60^{\circ}S$).
The model surface ozone in the MBL clearly shows
an interhemispheric asymmetry when compares ozone concentrations
at latitudes in the SH to the latitudes in the NH of the 
Atlantic and the Pacific, respectively.
For example, the averaged December 
ozone concentration over the Atlantic and the Pacific in the NH
is higher than in the SH (Figure~\ref{fig.8.1}(a)). 
This northward meridional gradient in ozone
concentration in December is reversed to the southward  
meridional gradient in July (Figure~\ref{fig.8.1}(b)),
indicating that the model meridional ozone
gradients in the MBL always point toward
the wintertime hemisphere. The change in the direction of 
concentration gradients closely follows the seasonal
movements.

The model predicts an area of high surface ozone concentrations
over the tropical south Atlantic, which extends across southern Africa,
the Indian Ocean, and Australia through September to October (Figure~\ref{fig.8.1}(c)). 
Based on satellite 
and ozonesonde measurements, {\it Jenkins et al.} [1997] reported 
that this area shows the highest tropospheric ozone concentrations
through September to October. This indicates that the model calculated 
surface ozone is consistent with the
measurements over the southern MBL. 
Notice that the seasonal tropospheric ozone maximum in the tropical south
Atlantic was first recognised from satellite observations by
{\it Fishman et al.} [1986, 1991], followed by many subsequent studies
[e.g., {\it Thompson et al.}, 1996; {\it Jacob et al.}, 1996; 
{\it Diab et al.}, 1996; {\it Browell et al.}, 1996].

\subsection{3.2. Comparison of Model with Surface Measurements}

\begin{figure*}[hp]
\vbox{
\vskip -0.0in
\centerline{
\leavevmode
\epsfxsize=6.0in
\epsfysize=3.0in
\epsfbox{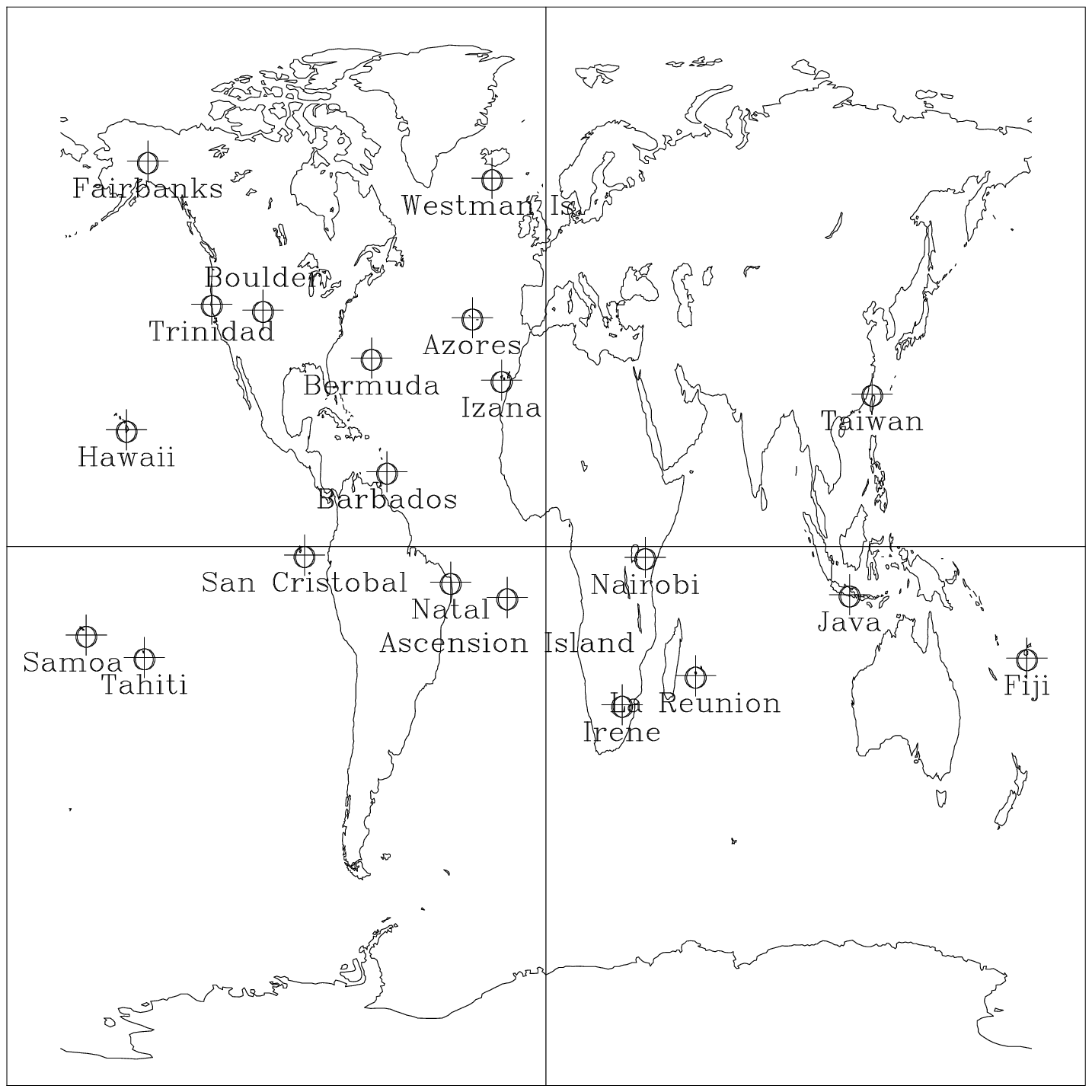}
}
}
\caption{ \label{fig.sites} Distribution of surface and ozonesonde
measurement sites used for this study.}
\end{figure*}

The results from IMS annual simulations were first compared with surface measurements
in the remote MBL environment. Figure~\ref{fig.sites} shows a
global distribution of the surface and ozonesonde measurement sites used for
the following comparison. The annual surface ozone measurements at
Westman Is., Bermuda, Mauna Loa, and Samoa
were taken from the NOAA 
CMDL Surface Ozone Data [{\it S.J. Oltmans}, 2001, personal communications; see also
{\it Oltmans and Levy}, 1994].
In addition to these sites, ozonesonde measurements were taken from the 
same CMDL source [{\it S.J. Oltmans}, 2001, personal communications], 
and the NASA SHADOZ data [{\it Thompson and Witte}, 1999].

%\subsubsection{3.2.1. Seasonal cycles over the MBL environments}

\begin{figure*}[hp]
\vbox{
\vskip -0.0in
\centerline{
\leavevmode
(a) 
\epsfxsize=3.0in
\epsfysize=1.5in
\epsfbox{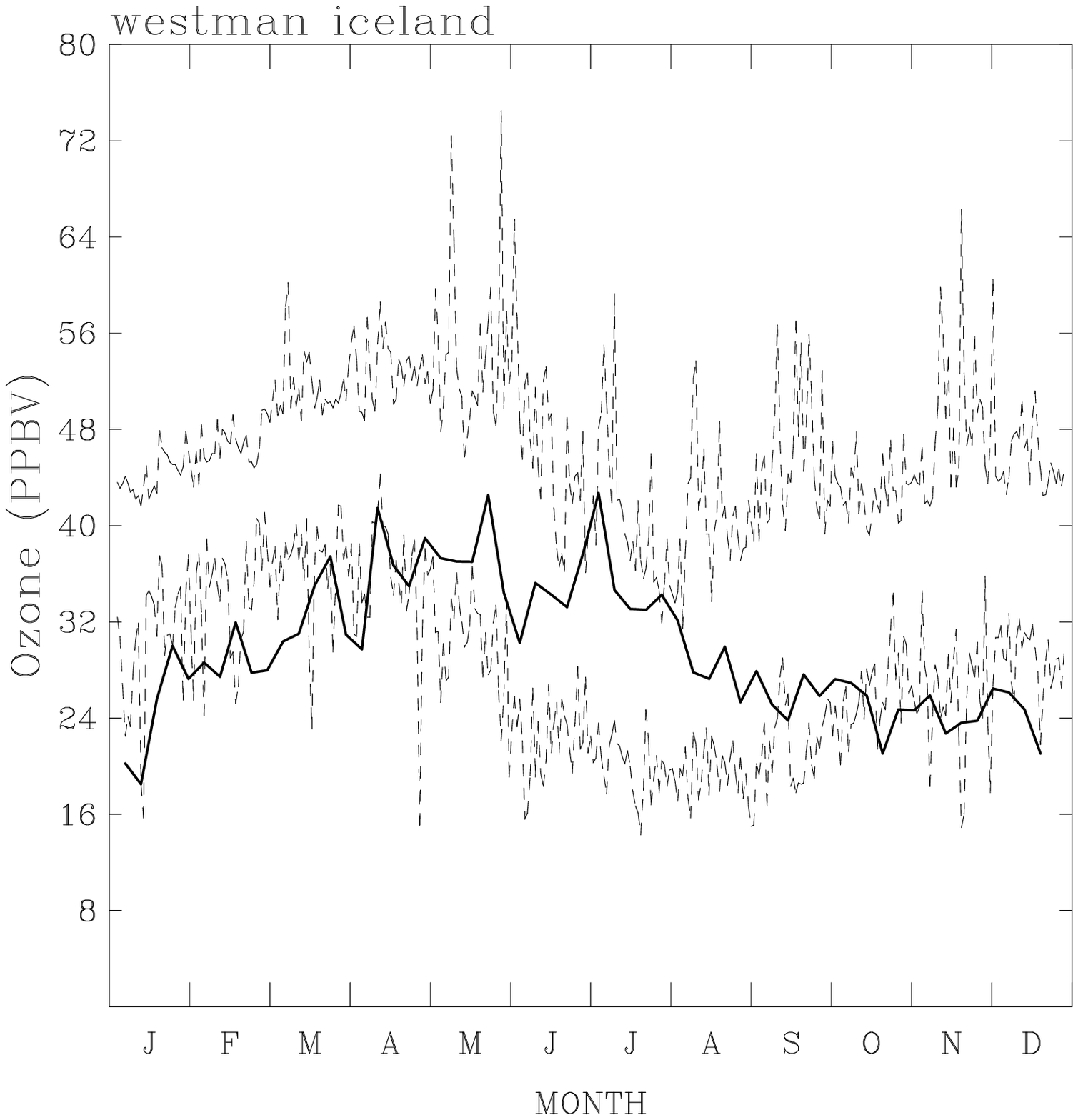}
(b)
\epsfxsize=3.0in
\epsfysize=1.5in
\epsfbox{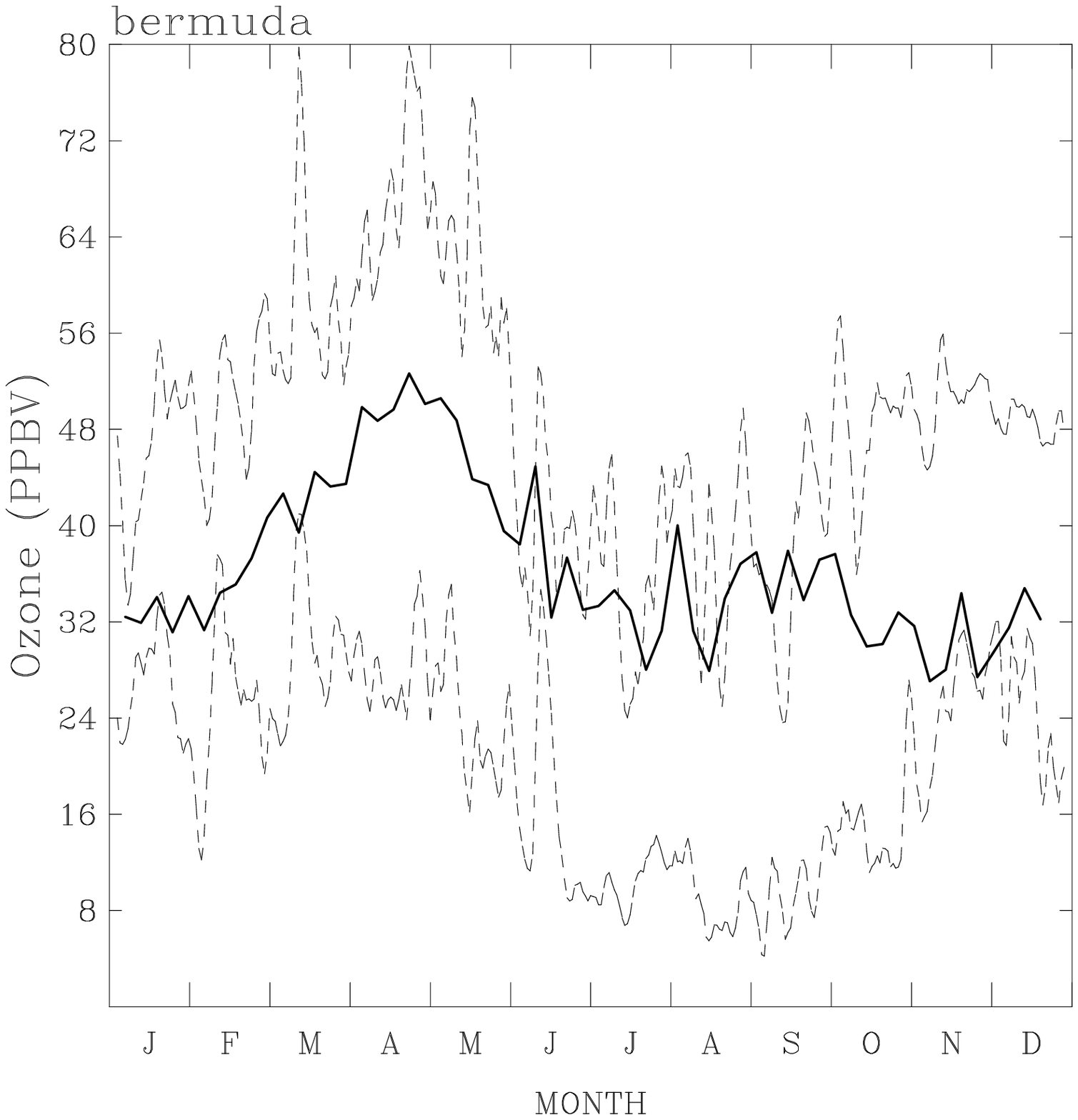}
}
\vskip -0.0in
\centerline{
\leavevmode
(c) 
\epsfxsize=3.0in
\epsfysize=1.5in
\epsfbox{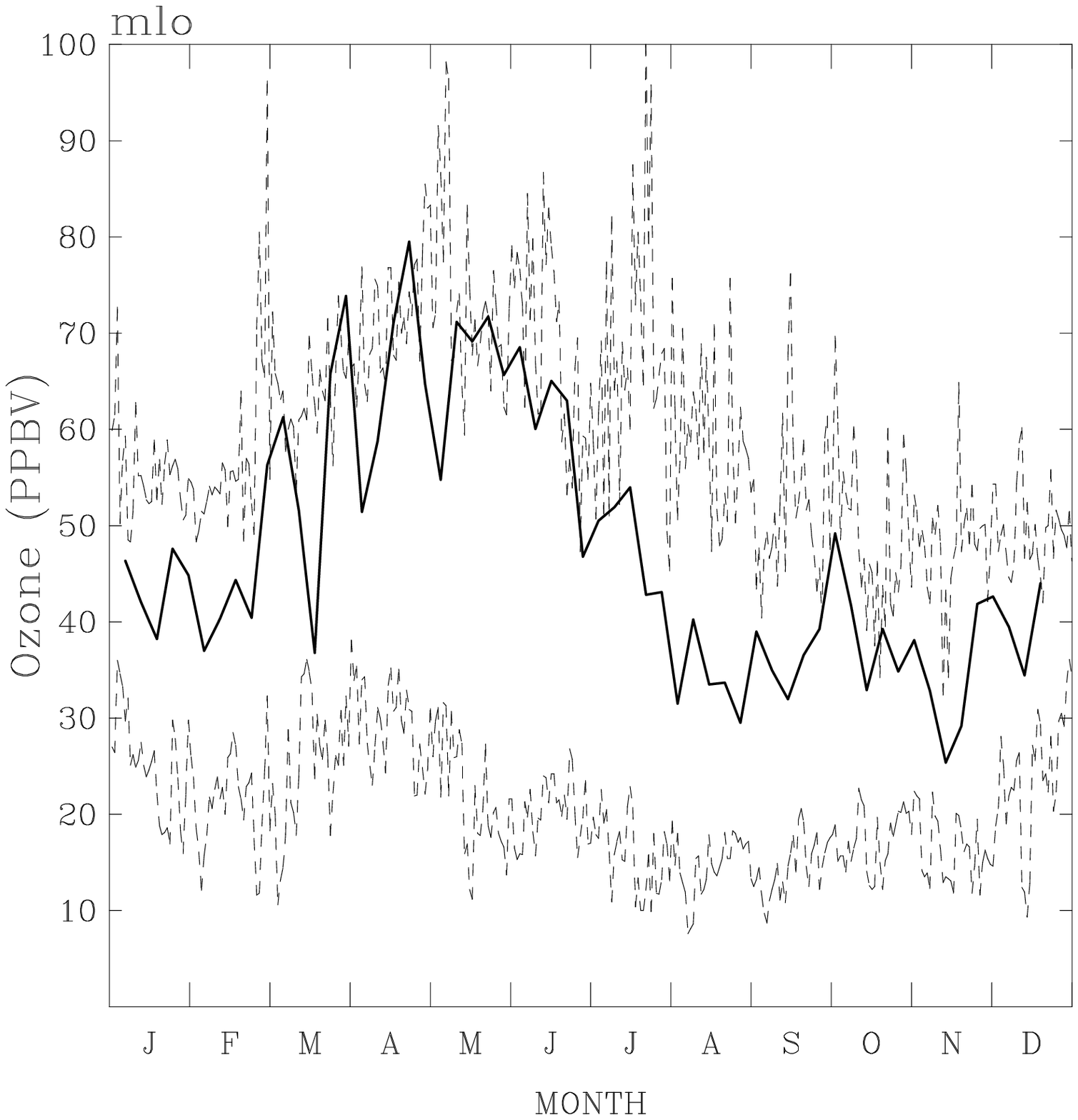}
(d)
\epsfxsize=3.0in
\epsfysize=1.5in
\epsfbox{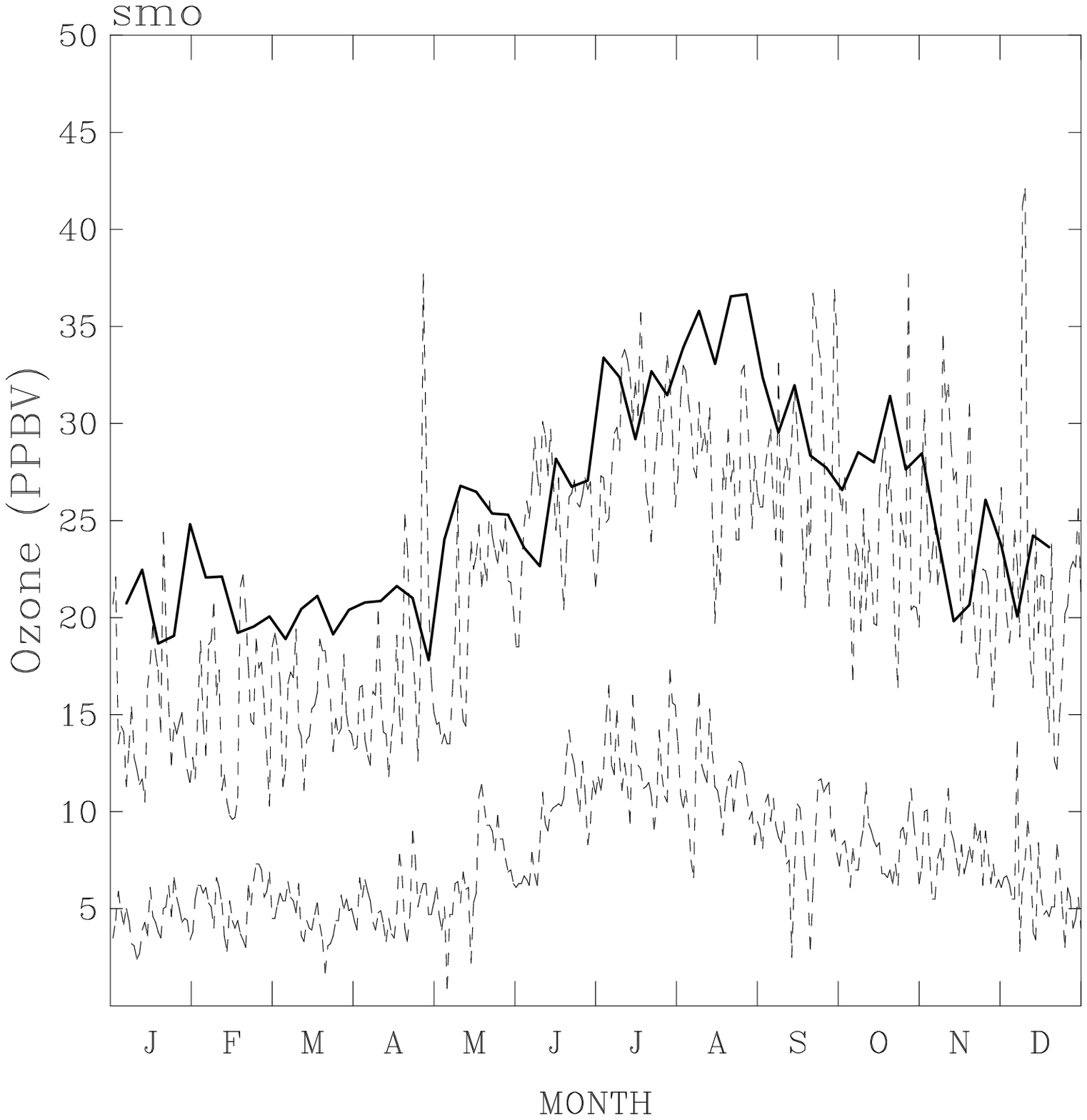}
}
}
\caption{ \label{fig.8.14.1} Comparison of modelled (solid thick lines) 
seasonal cycles  
of $O_3$ (ppbv) at (a) Westman, Iceland, (b) Bermuda, (c) Mauna Loa, 
and (d) Samoa with the measurements (thin dashed lines).
Two measured $O_3$ levels (for 1988-1992, except at Westman where the 1992-1997 data were used) 
are shown here, one for the daily maximum,
while the other one for the daily minimum.}
\end{figure*}

Figure~\ref{fig.8.14.1} shows time-series plots of modelled and 
observed $O_3$ levels at four sites located in the 
remote MBL. We compare model results with
two measured $O_3$ levels (for the 1988-1992), one for the daily maximum, and 
the other one for the daily minimum. While the observed daily minima represent
the indigenous local background 'clean' conditions, the observed daily maxima
clearly indicate the nonindigenous influences such as elevated $O_3$ from the
upper troposphere, or due to the long-range transport of high $O_3$ and 
anthropogenic $O_3$ precursors from industrial and biomass burning areas.
These time-series plots show very distinct and easily identifiable
spring ozone maxima at these MBL sites. While it has long been
observed that the spring maximum is a NH phenomenon [e.g., {\it Monks}, 2000],
the SH observations at Samoa also shows very distinct spring time
maximum comparable to those NH sites. 

The observed 
spring ozone behaviour is generally well reproduced by the
model at these sites, and the modelled  
ozone levels generally fall within 
the observed ranges at Westman, Bermuda, and Mauna Loa, 
and is at the upper
bound of the observed ozone concentrations at Samoa. 
Though the model overestimates ozone at some tropical MBL locations,
the observed seasonal cycles are closely reproduced by the model. This
indicates that the model is capturing the correct sense of  
the processes controlling ozone variation at
remote MBL environments.

Notice that the time of observed minimum (JJA) is not reproduced by the model
at Westman. This indicates that model underestimates the processes contributing to the
ozone levels at this location in other seasons. For example, in the simulation without
considering tropospheric emissions (see part 3),
the model can only produce ozone close to the daily minimum measurements. Hence,
the background ozone concentration at higher latitudes is likely to be too low
in the seasons through autumn to late winter.
The model also overestimates ozone at
Samoa, indicating that too much ozone has been transported/produced in the southern MBL.

We note that the observed daily maxima are 2 to 4 times that
of the daily minima, indicating that the MBL environment is actually
very sensitive to the air from elsewhere such as the upper troposphere 
and industrial and biomass burning areas.
%
% More sensitive to transport than to photochemistry
%
These transport-driven sensitivities are much higher than those
driven by the photochemical loss process, which is on the order 
%
% What is the magnitude of diurnal variation of MBL O3? 
%
of no more than a few ppbv per day
[e.g., {\it Paluch et al.}, 1994; {\it Monks et al.}, 2000]. 
This indicates that ozone in the MBL at these sites are largely dominated by
the processes such as long-range          
transport of ozone-riched air, 
cloud convective transport, mixing and dry deposition in the MBL,
and cross-tropopause
transport.

\subsection{3.3. Comparison of Model with Ozonesondes}

While the previous surface comparisons show pronounced seasonal cycles and distinctive
interhemispheric asymmetry in $O_3$ in the low latitude MBL, 
these comparisons were limited to the surface [e.g., {\it Oltmans et al.}, 1996].
Many components such as free tropospheric $O_3$ and other $O_3$ precursor distribution, 
$O_3$ exchange in the upper troposphere and lower stratosphere, and atmospheric transport
processes are crucial for understanding the chemical behaviour over the remote marine 
troposphere. In this section we compare modelled ozone vertical profiles with ozonesonde
measurements
taken from the NASA SHADOZ data and the NOAA CMDL ozone data for the period 1998-1999.

\subsubsection{3.3.1. The Atlantic} 

\begin{figure*}[hp]
\vbox{
\vskip -0.0in
\centerline{
\leavevmode
(a)
\epsfxsize=1.5in
\epsfysize=3.0in
\rotatebox{-90.}{\epsfbox{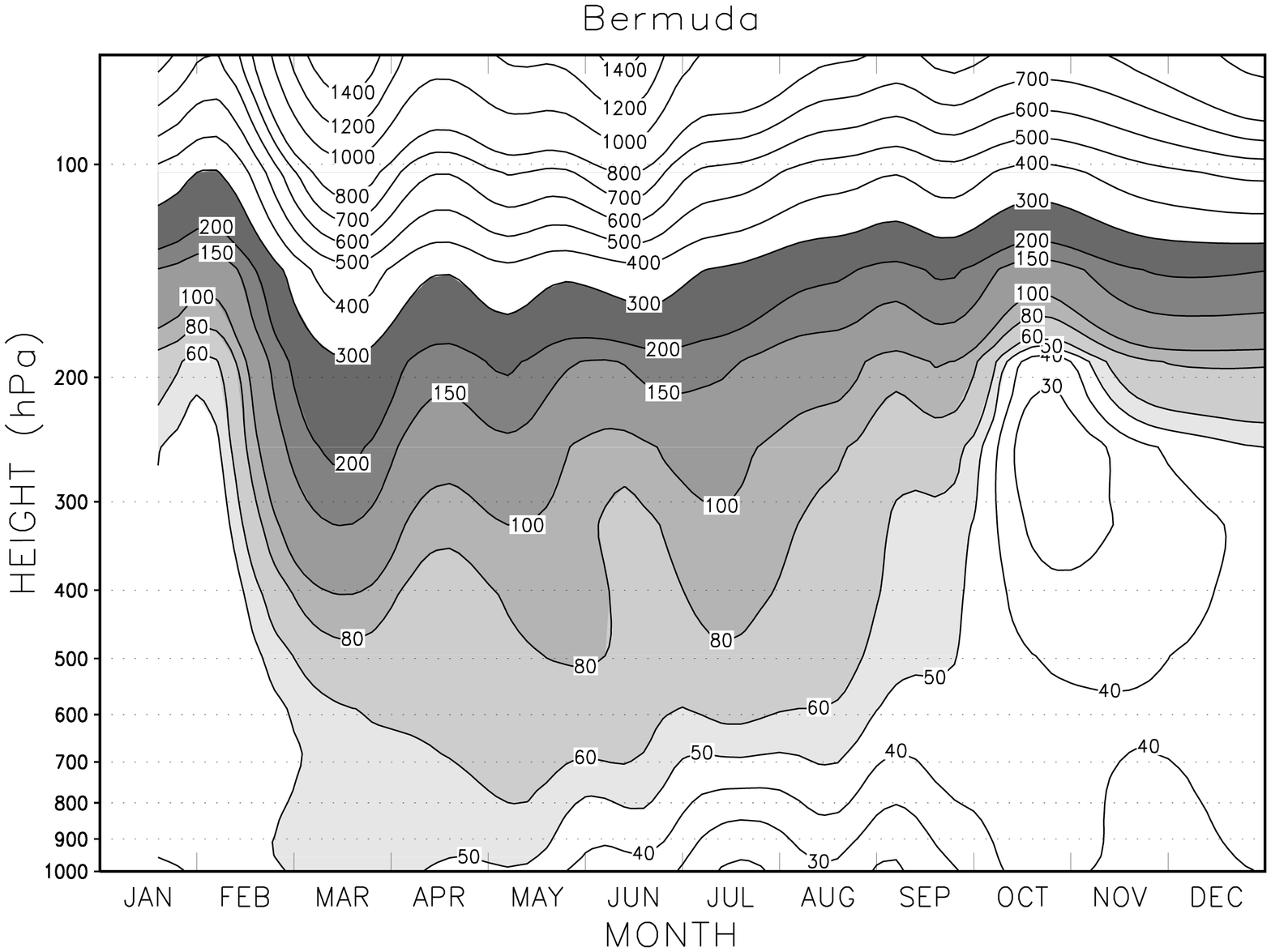}}
(b)
\epsfxsize=1.5in
\epsfysize=3.0in
\rotatebox{-90.}{\epsfbox{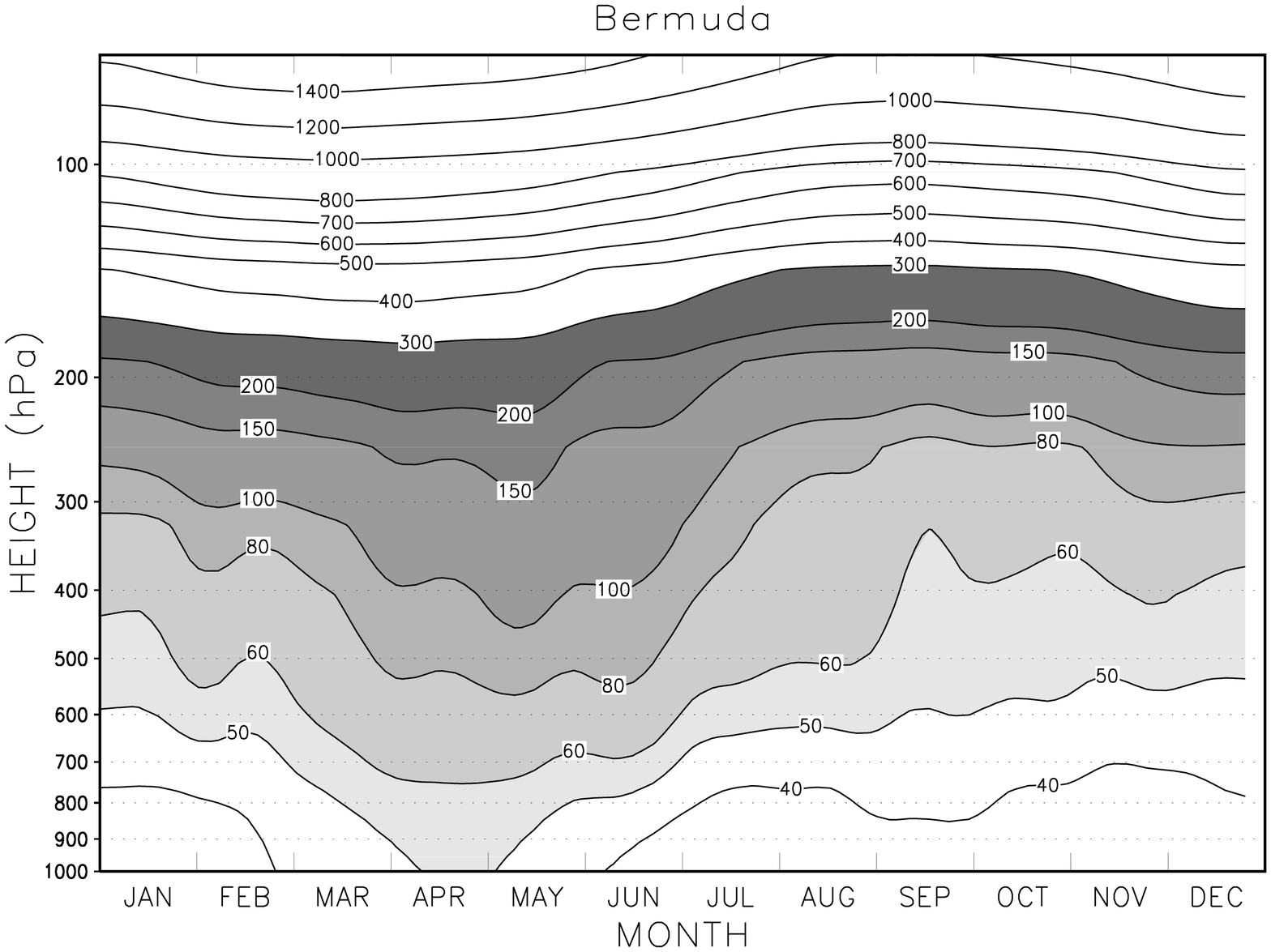}}
}
\vskip -0.0in
\centerline{
\leavevmode
(c)
\epsfxsize=1.5in
\epsfysize=3.0in
\rotatebox{-90.}{\epsfbox{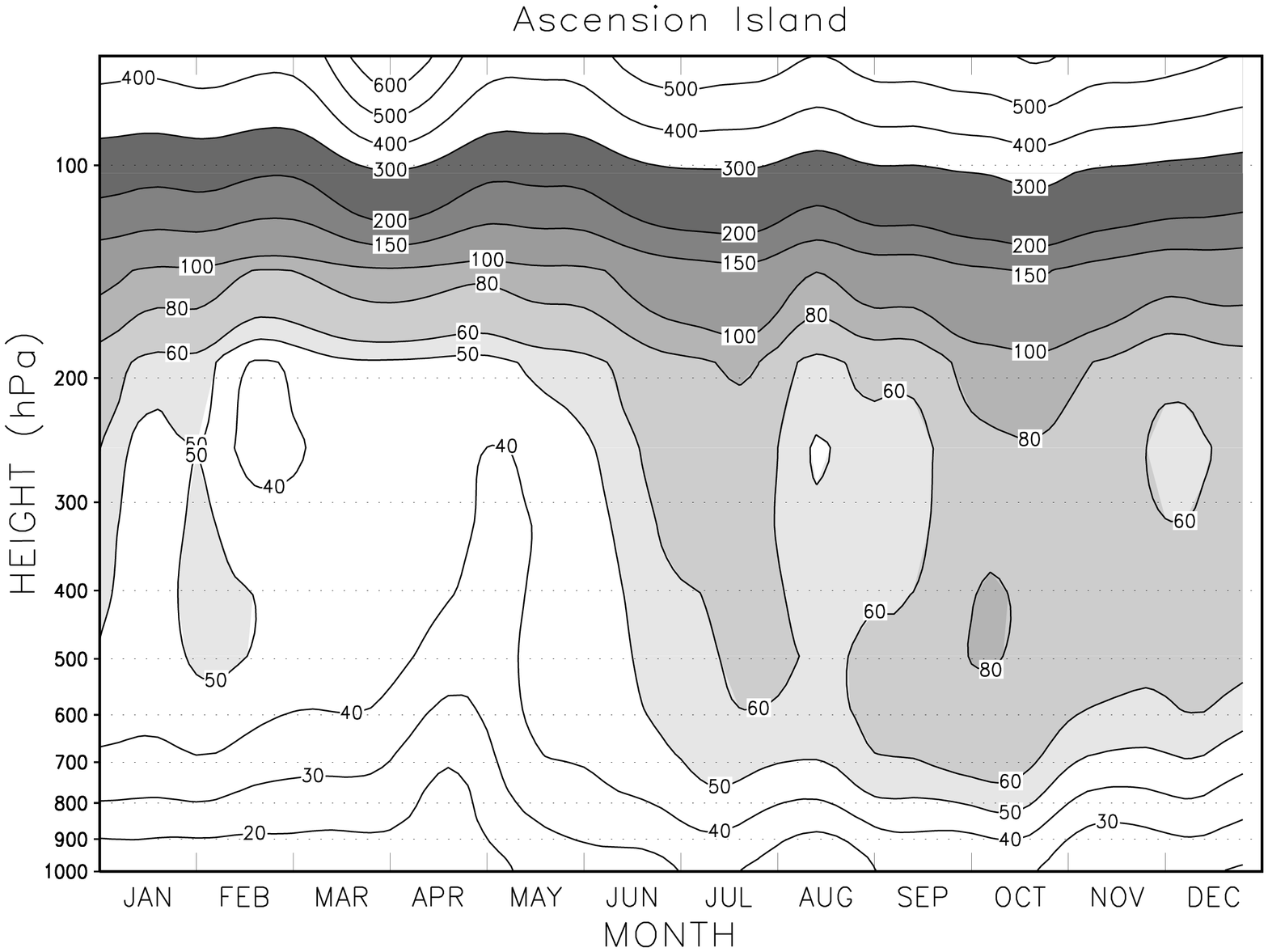}}
(d)
\epsfxsize=1.5in
\epsfysize=3.0in
\rotatebox{-90.}{\epsfbox{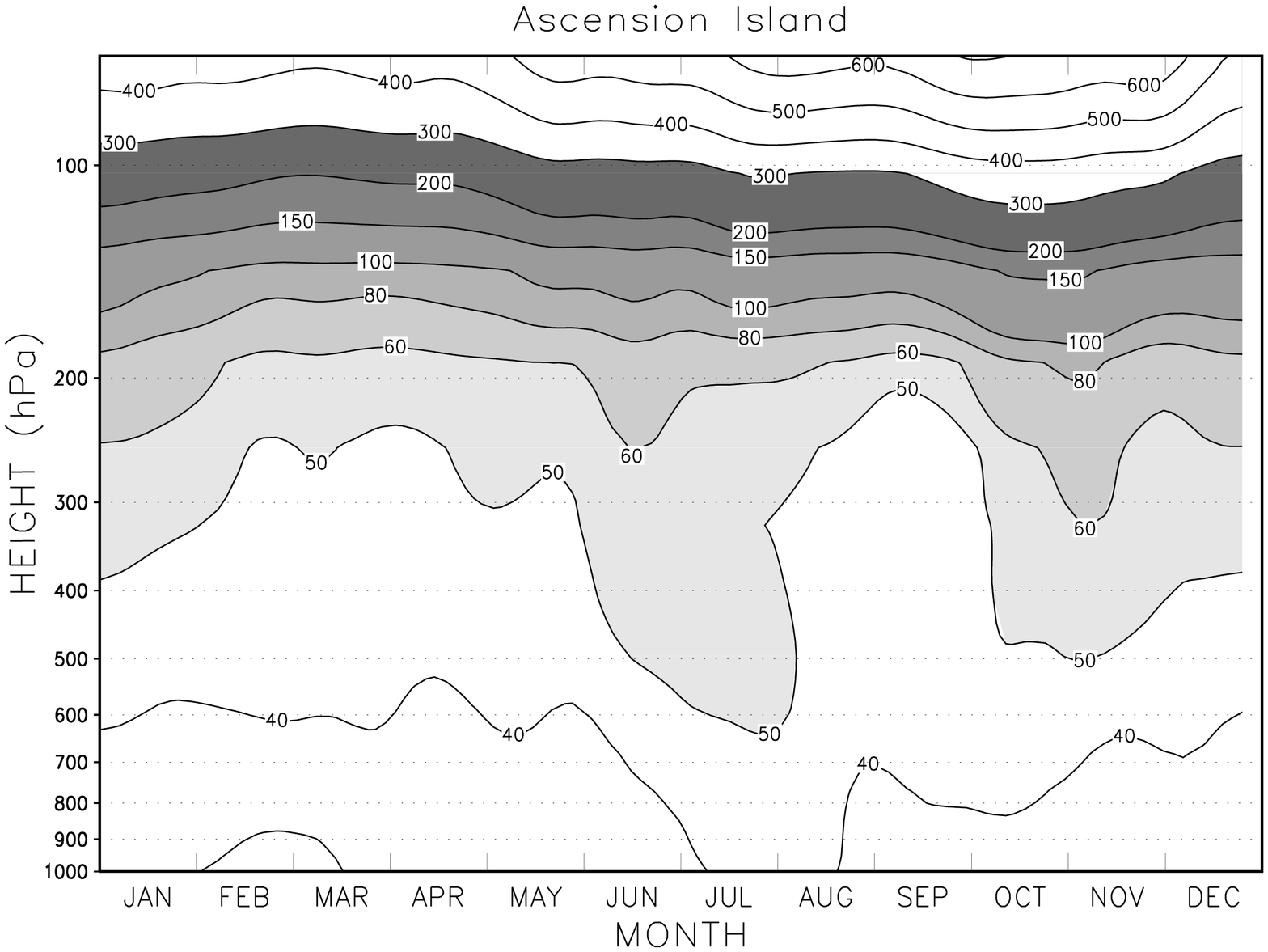}}
}
}
\caption{ \label{fig.ozonesonde.at} Time-height cross sections of $O_3$ (ppbv) from measurements
at Bermuda ($32^{\circ}N$, $65{\circ}W$) (a) and Ascension Island ($8^{\circ}S$, $14^{\circ}W$) (c). 
The model calculation for
these locations are shown in (b) and (d), respectively.}
\end{figure*}

Figure~\ref{fig.ozonesonde.at} shows time-height cross sections of
measured and modelled $O_3$ at northern (Bermuda) and 
southern (Ascension Island) Atlantic.
%
% Measurements: Northern Atlantic
%
For the northern Atlantic site (Figure~\ref{fig.ozonesonde.at}(a)),
analyses of one year of vertical soundings of ozone
show that high values of ozone extend from the upper troposphere to the middle and lower
troposphere during the NH spring (March-May). High levels of ozone are also seen
in the lower stratosphere during this period. 
%
% Models: Northern Atlantic
%
The model calculated annual variations of ozone in the troposphere at this location
is shown in Figure~\ref{fig.ozonesonde.at}(b). The observed high ozone concentrations
from the upper to the middle and lower troposphere during the NH spring are generally 
well reproduced by the model. 
%
% Extend downward from the tropopause and extend upward from the surface 
%
Both model and ozonesondes show a consistent picture of the ozone distribution in the troposphere 
during the NH spring compared with other seasons: Larger ozone concentrations extend downward
from the tropopause to near the surface, while smaller ozone concentrations
extend upward from the surface to near the tropopause. These characteristics are consistent
with other analyses [{\it Oltmans et al.}, 1996].

%
% What others said about Northern Atlantic: Stratosphere-dominated
%
{\it Moody et al.} [1995] suggested that the elevated ozone concentrations in the
midtroposphere at this location during this period is associated with downward 
transport of ozone from
the upper and lower stratosphere.
Based on the analyses of summer and spring ozonesondes at five locations over the  
North Atlantic, {\it Oltmans et al.} [1996] found the connection between large
ozone mixing rations and dry air in the middle and upper troposphere with large
ozone values in the tropopause region. They suggested that the stratosphere plays   
a major role in loading the troposphere with ozone, and high ozone events usually
extend downward from the tropopause region. From trajectory analyses showing 
the history of transport path, {\it Oltmans et al.} [1996] concluded that 
the upper troposphere and lower stratosphere was the source of elevated 
ozone in the troposphere.

%
% Measurements: Southern Atlantic 
%
For the southern Atlantic site (Figure~\ref{fig.ozonesonde.at}(c)),
high ozone concentrations
are seen in the
troposphere
during the SH spring (September-November).
The spread of high ozone levels extends from the upper troposphere downward to 
near the surface during this season. 
%
% Models: Southern Atlantic
%
Model calculations (Figure~\ref{fig.ozonesonde.at}(d)) at this southern
Atlantic site show similar spring ozone maxima extending from the upper troposphere
downward, in accordance with the measurements, though the modelled high ozone in the upper troposphere 
does not extend as far down as in the measurements. 

%
% What others said about Southern Atlantic: Biomass burning 
%
The enhanced tropospheric ozone, which occurs between July and October and observed at
Ascension Island, is linked to dry season biomass burning [{\it Diab et al.}, 1996]. 
During September-October, gases from extensive fires in Brazil were transported by convective
storms into the upper troposphere where tropospheric ozone was photochemically produced
and advected eastward over the south Atlantic, while the widespread fires in 
the deep convection-free central Africa were advected at low altitudes over the Atlantic
[{\it Browell et al.}, 1996; {\it Jacob et al.}, 1996; {\it Thompson et al.}, 1996].
The lower modelled ozone concentration in the troposphere compared with the ozonesondes indicate
that biomass burning emissions in the equatorial south America and central Africa 
are too low during the biomass burning season.

%
% Interhemispheric asymmetry: clearly seen for the measurements, and
%                             well reproduced by the model
%

The above analyses show an interhemispheric asymmetry in the tropospheric ozone distribution
over the Atlantic basin. More persistent and widespread 
ozone maxima were observed and modelled during the hemispheric spring months, 
from March to May for the NH, and
from September to November for the SH, than any other seasons of the year. 
Though the season (hemispheric spring) is the same, the sources of elevated
ozone in the troposphere are different when comparing the NH with the SH.
For the NH Atlantic basin, the downward transport of high ozone from upper troposphere and lower
stratosphere are likely to be the major sources. For the SH Atlantic basic, the biomass burning  
emissions from continents, together with following cloud convective transport and/or large
scale transport are likely to be the major contributors (see following part 2 and part 3
papers for further discussions).

\subsubsection{3.3.2. Western Indian Ocean} 
\begin{figure*}[hp]
\vbox{
\vskip -0.0in
\centerline{
\leavevmode
(a)
\epsfxsize=1.5in
\epsfysize=3.0in
\rotatebox{-90.}{\epsfbox{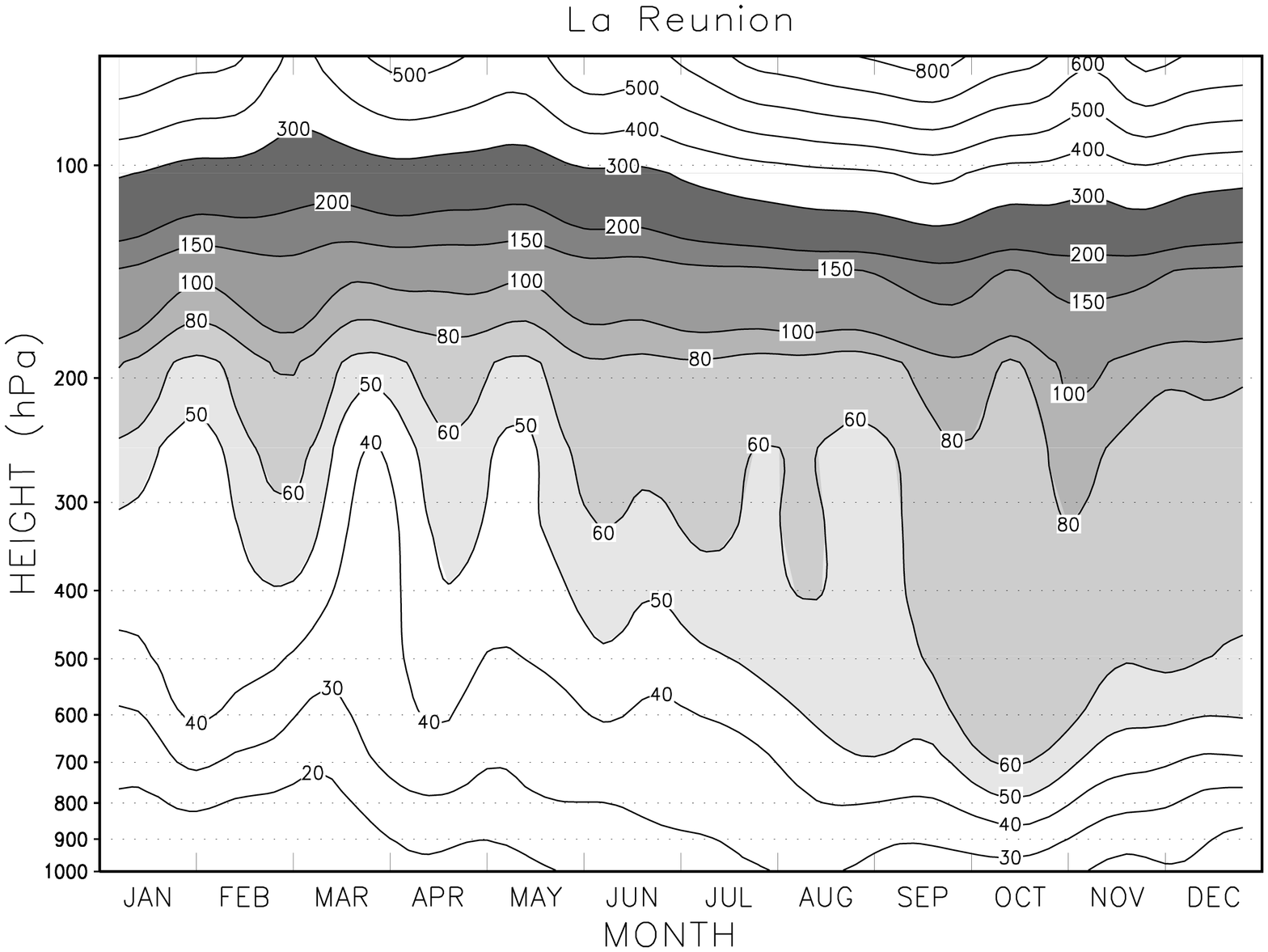}}
(b)
\epsfxsize=1.5in
\epsfysize=3.0in
\rotatebox{-90.}{\epsfbox{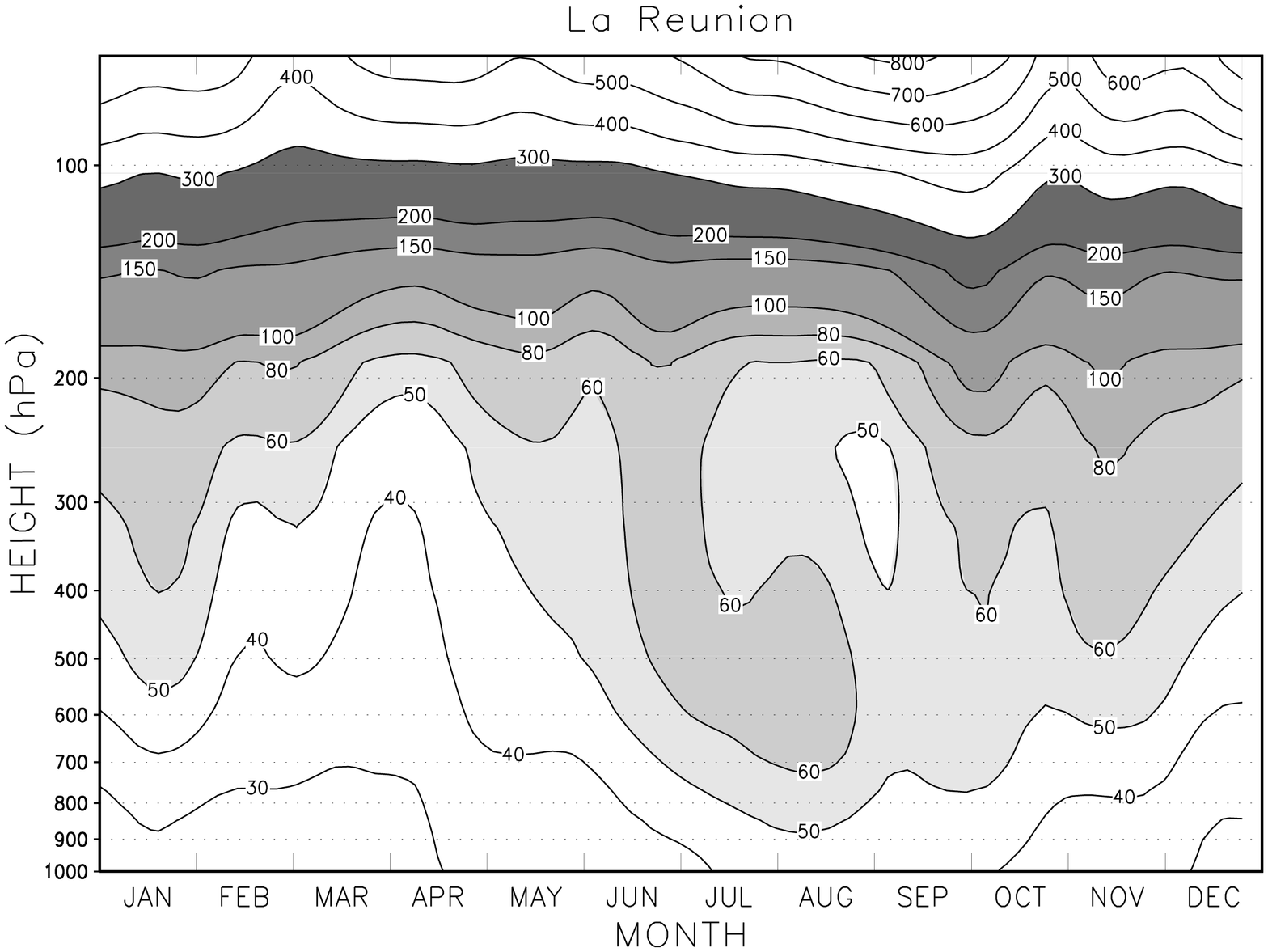}}
}
\vskip -0.0in
\centerline{
\leavevmode
(c)
\epsfxsize=1.5in
\epsfysize=3.0in
\rotatebox{-90.}{\epsfbox{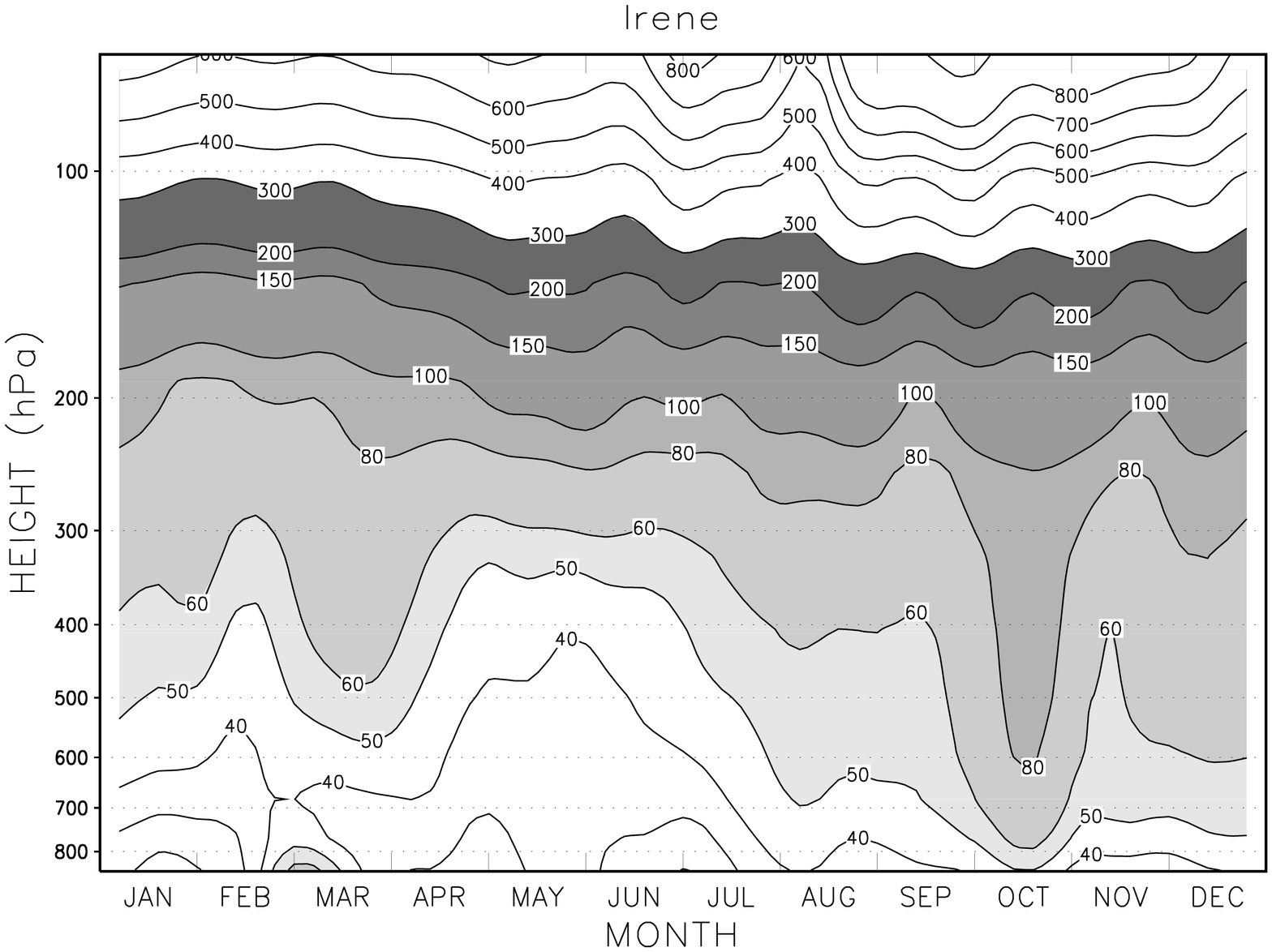}}
(d)
\epsfxsize=1.5in
\epsfysize=3.0in
\rotatebox{-90.}{\epsfbox{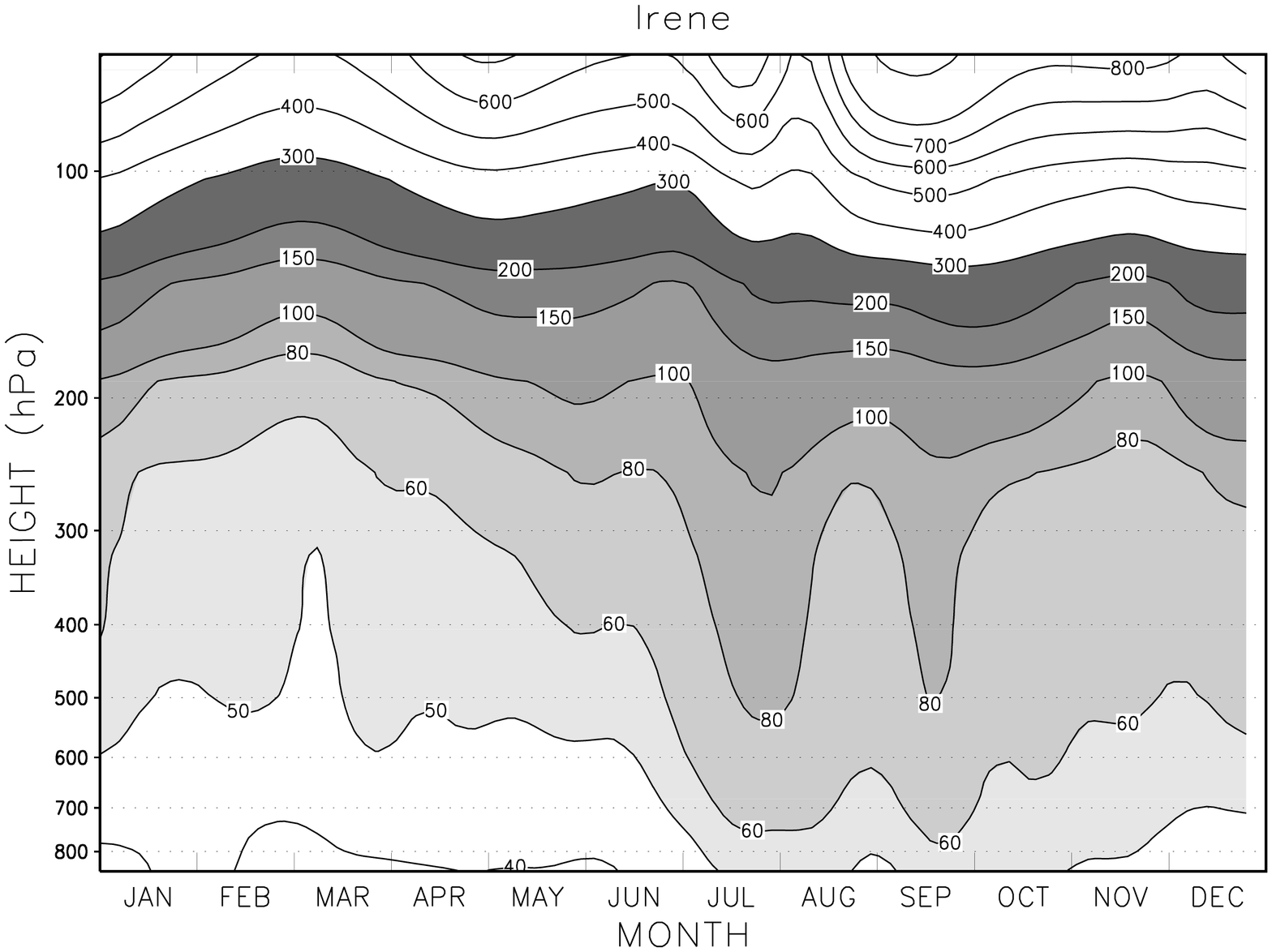}}
}
}
\caption{ \label{fig.ozonesonde.in} Time-height cross sections of $O_3$ (ppbv) from measurements at 
La Reunion ($21^{\circ}S$, $56^{\circ}E$) (a) and Irene ($26^{\circ}S$, $28^{\circ}E$) (c). 
The model calculation for
these locations are shown in (b) and (d), respectively.}
\end{figure*}

Figure~\ref{fig.ozonesonde.in} shows comparisons of ozone profiles from ozonesondes with
model calculations at two locations over the SH western Indian Ocean. 
Extensive high ozone concentrations from the lower to upper troposphere are seen at these sites.
Model calculations show similar SH spring maxima in the
troposphere compared with the measurements. 
%
% Reunion Island -- T2 (biomass burning)
%
Both locations experienced the same sources for the elevated spring ozone 
in the troposphere as to the SH Atlantic sites of Ascension Island and
Natal [{\it Diab et al.}, 1996; {\it Baldy et al.}, 1996]. 
Analysis of one-year ozonesondes at Reunion Island (Figure~\ref{fig.ozonesonde.in}(c)) shows high 
levels of ozone are observed in the lower to upper troposphere during the SH spring 
(September-November). {\it Baldy et al.} [1996] reported that the elevated ozone in the
free troposphere during this period of time at this island is  
concomitant with active biomass burning in the southeastern African
continent and Madagascar. 
{\it Thompson et al.} [1996] reported that features of elevated tropospheric $O_3$ ($\ge$ 90 ppbv)
extend in a band from $0^{\circ}$ to $25^{\circ}S$, over SE Indian Ocean, Africa, the Atlantic,
and eastern South America during September-October. They showed a strong connection between
regions of high ozone and concentrated biomass burning. 

\subsubsection{3.3.3. Western Pacific Ocean} 

\begin{figure*}[hp]
\vbox{
\vskip -0.0in
\centerline{
\leavevmode
(a)
\epsfxsize=1.5in
\epsfysize=3.0in
\rotatebox{-90.}{\epsfbox{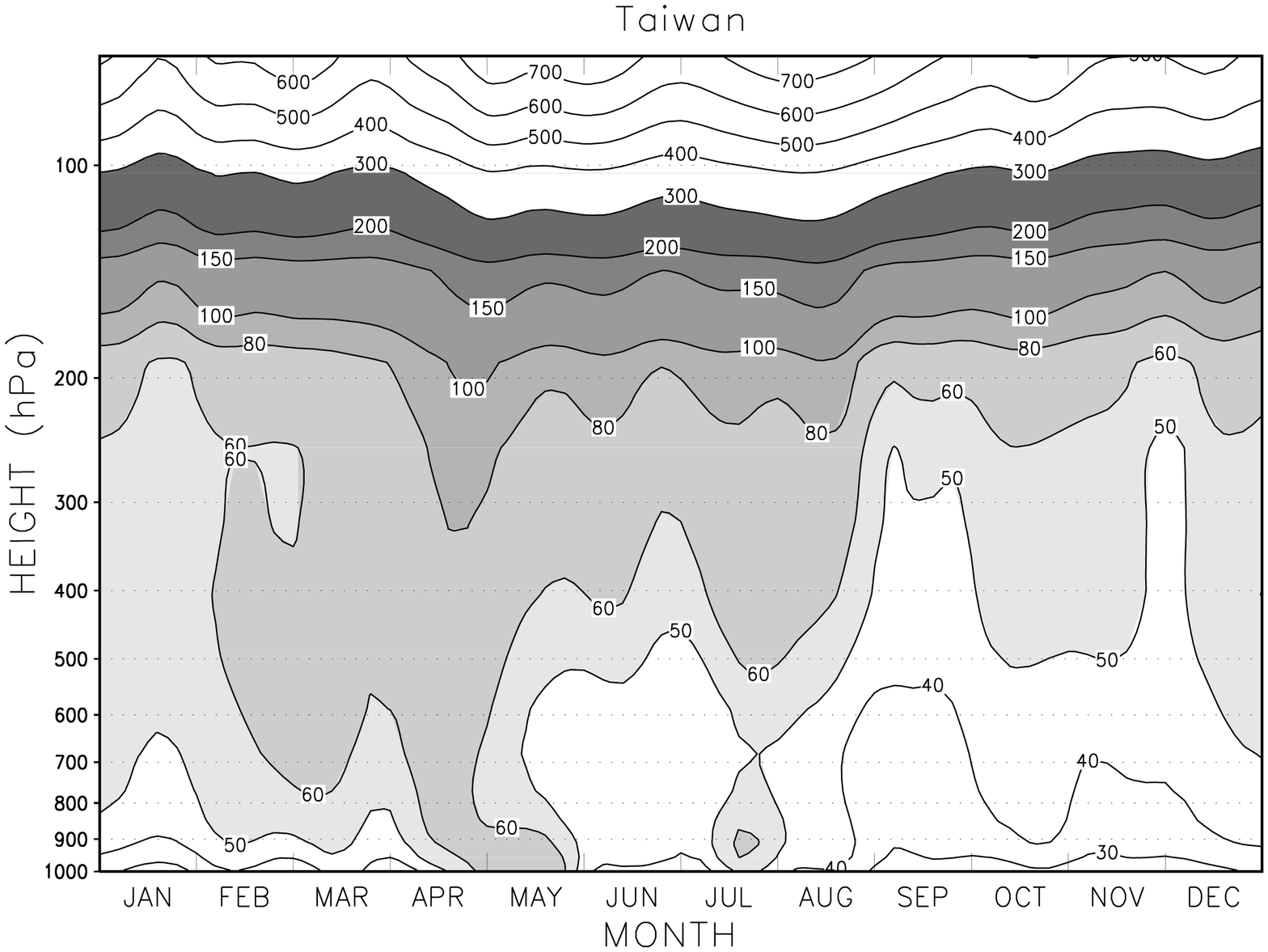}}
(b)
\epsfxsize=1.5in
\epsfysize=3.0in
\rotatebox{-90.}{\epsfbox{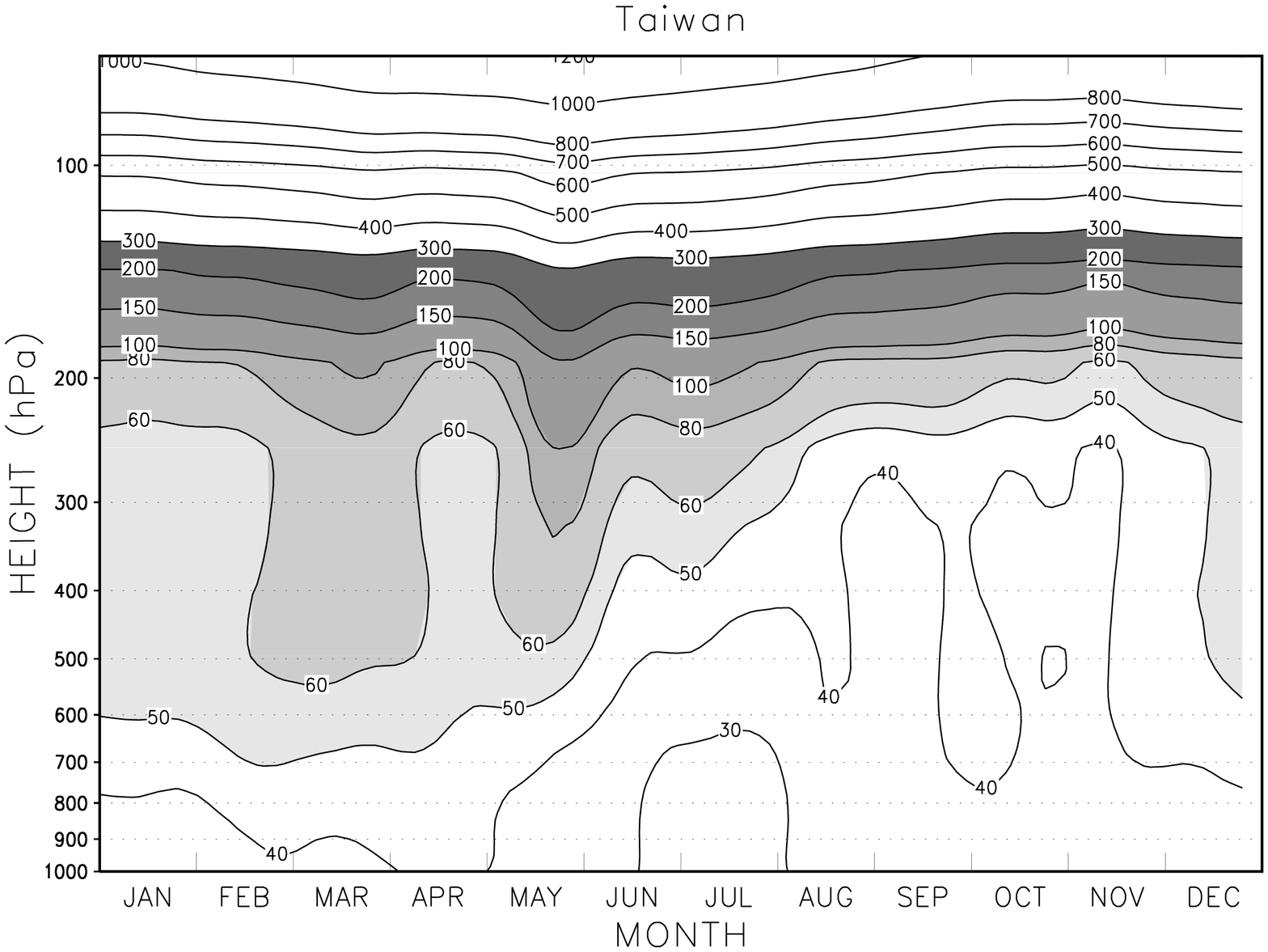}}
}
\vskip -0.0in
\centerline{
\leavevmode
(c)
\epsfxsize=1.5in
\epsfysize=3.0in
\rotatebox{-90.}{\epsfbox{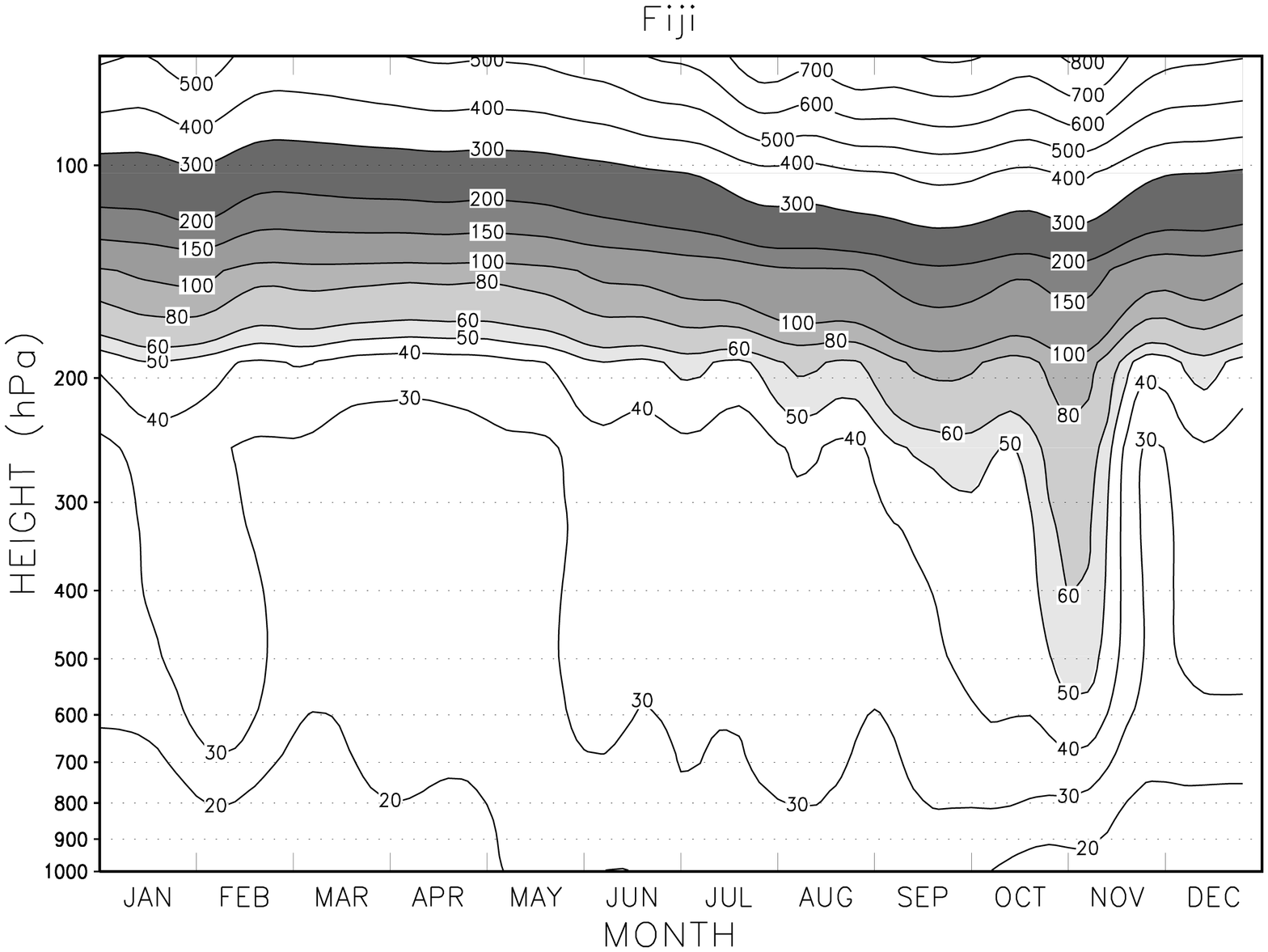}}
(d)
\epsfxsize=1.5in
\epsfysize=3.0in
\rotatebox{-90.}{\epsfbox{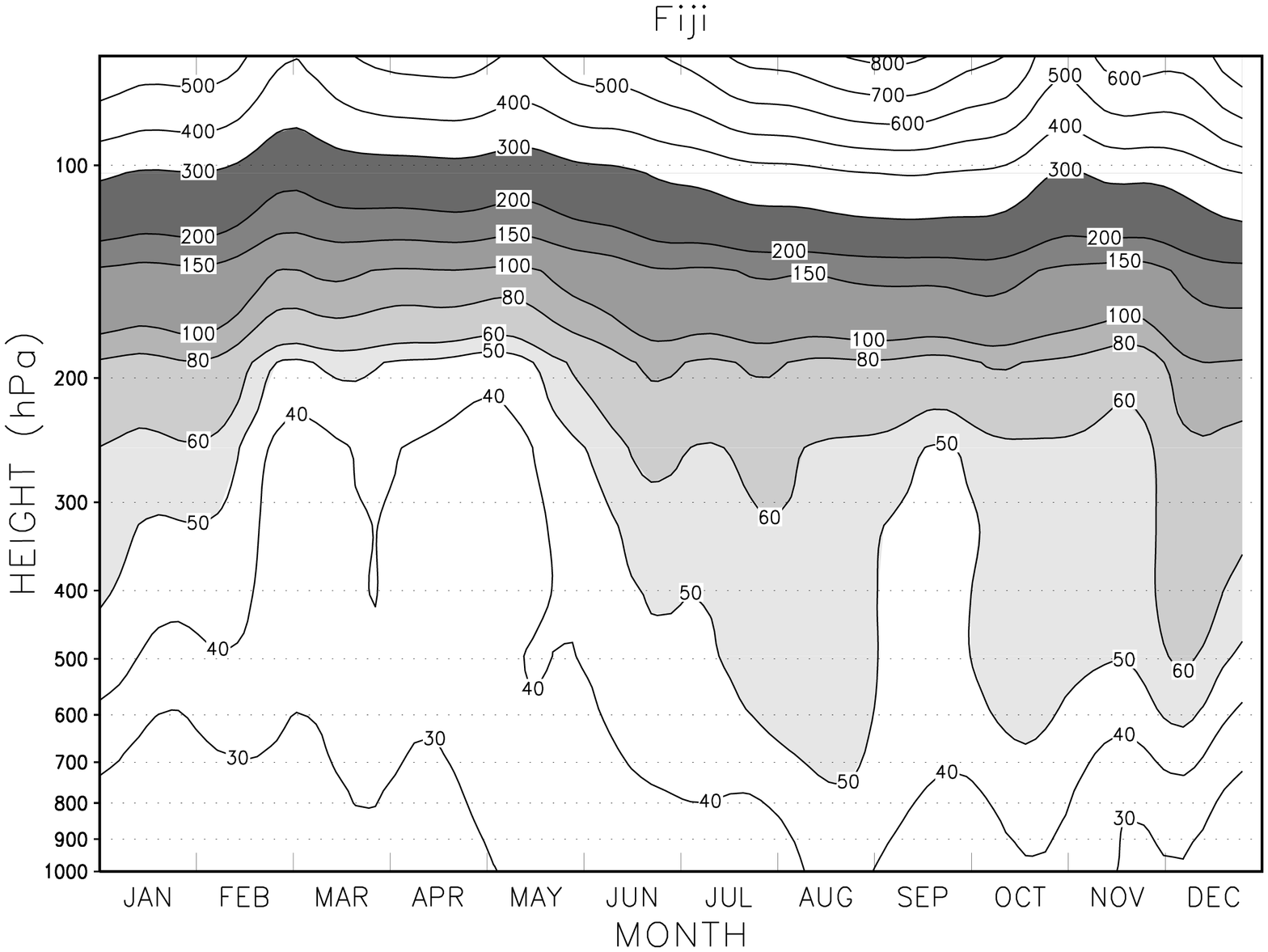}}
}
}
\caption{ \label{fig.ozonesonde.wpa} Time-height cross sections of $O_3$ (ppbv) from measurements at  
Taiwan ($25^{\circ}N$, $121^{\circ}E$) (a) and Fiji ($17^{\circ}S$, $179^{\circ}E$) (c). 
The model calculation for
these locations are shown in (b) and (d), respectively.}
\end{figure*}

The observations and our previous modelling at sites in the southern Atlantic 
basin showed that
large-scale biomass burning emissions provide sources of elevated ozone over the 
tropical south Atlantic. Long-range transport of biomass burning pollution
could affect ozone on a hemispheric scale [{\it Fishman et al.}, 1991; {\it Schultz et al.}, 1999]. 
Here we examine the extent of seasonal biomass burning influences over the Pacific basin.
Figure~\ref{fig.ozonesonde.wpa} show time-height cross sections of vertical ozone profiles
form ozonesondes and model calculations at Taiwan (in the subtropical western North Pacific), 
and Fiji (in the subtropical western South Pacific).

Observed ozone profiles at Taiwan shows similar spring ozone maxima in the troposphere as to
the sites in the northern Atlantic, and Hawaii in the eastern North Pacific (see next section).
The model shows a similar spring ozone maximum, though the elevated ozone concentrations do not 
extend as far down to near the surface as in the measurements. The major differences at 
altitudes below 4 km ($\sim$ 600 hPa) coincide with the altitude range most influenced
by continental outflow [e.g., {\it Kajii et al.}, 1997; {\it Crawford et al.}, 1997]. 
This indicates that the model underestimates the impact of continental outflow of $O_3$ precursors
such as NO and NMHC [{\it Crawford et al.}, 1997]. 

The model calculations at Fiji show that elevated
tropospheric ozone occurs in August, and from October onward. 
The elevated ozone ($\ge$ 40 ppbv) calculated by the model at Java are seen from May to
August, instead of measured maxima from October to November.
{\it Schultz et al.} [1999] reported the importance of biomass burning emissions in South America
and Africa for the ozone budget at higher altitudes,
and the NOx decomposed from transported PAN at altitudes below 4 km over the tropical South Pacific.
This indicates that the model underestimates the impact from biomass burning 
emissions and the long-range transport of 
reactive nitrogen at Java, while overestimating the impact of biomass burning emissions at Fiji. 

\subsubsection{3.3.5. Central to Eastern Pacific} 

\begin{figure*}[hp]
\vbox{
\vskip -0.0in
\centerline{
\leavevmode
(a)
\epsfxsize=1.5in
\epsfysize=3.0in
\rotatebox{-90.}{\epsfbox{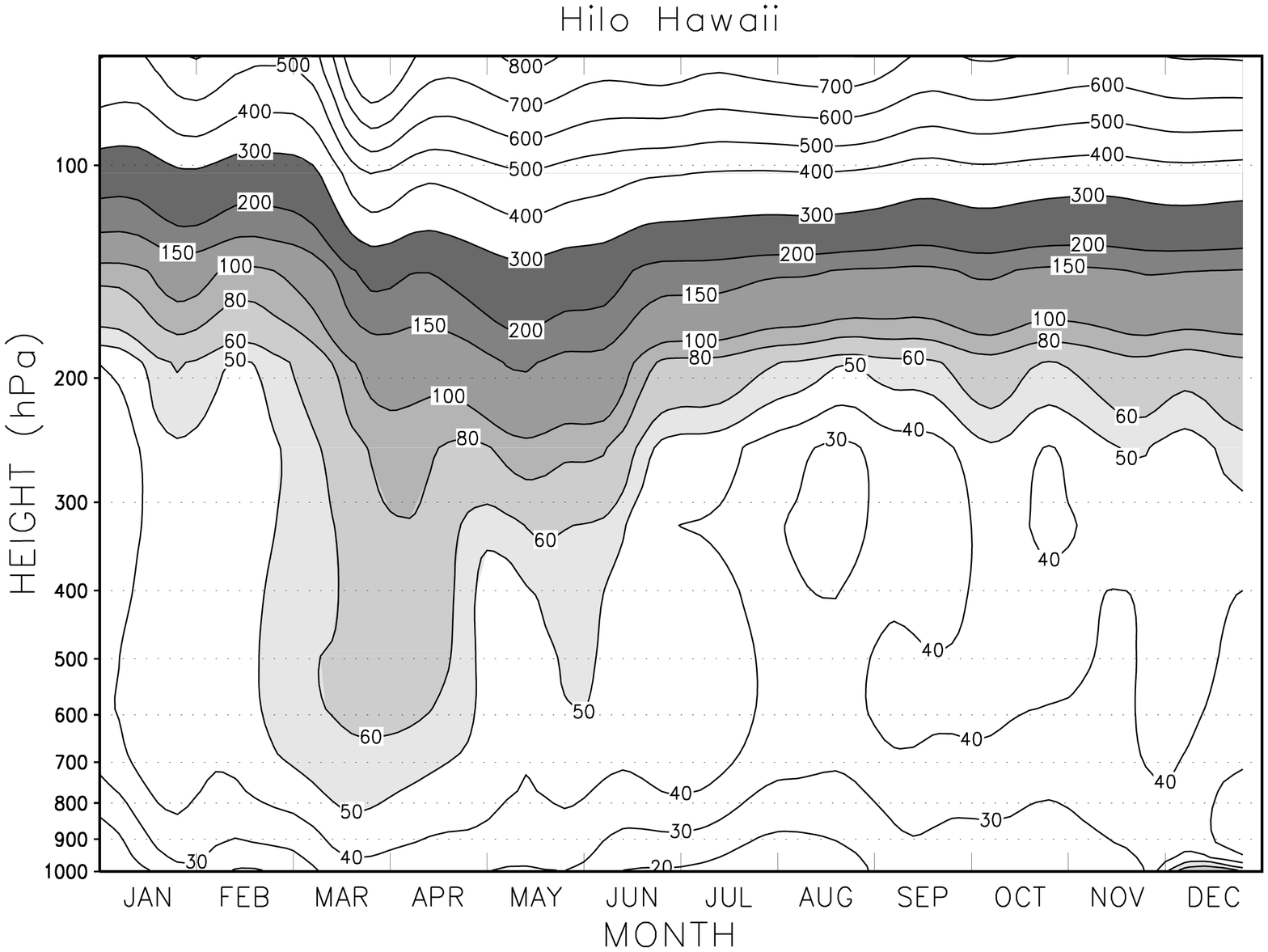}}
(b)
\epsfxsize=1.5in
\epsfysize=3.0in
\rotatebox{-90.}{\epsfbox{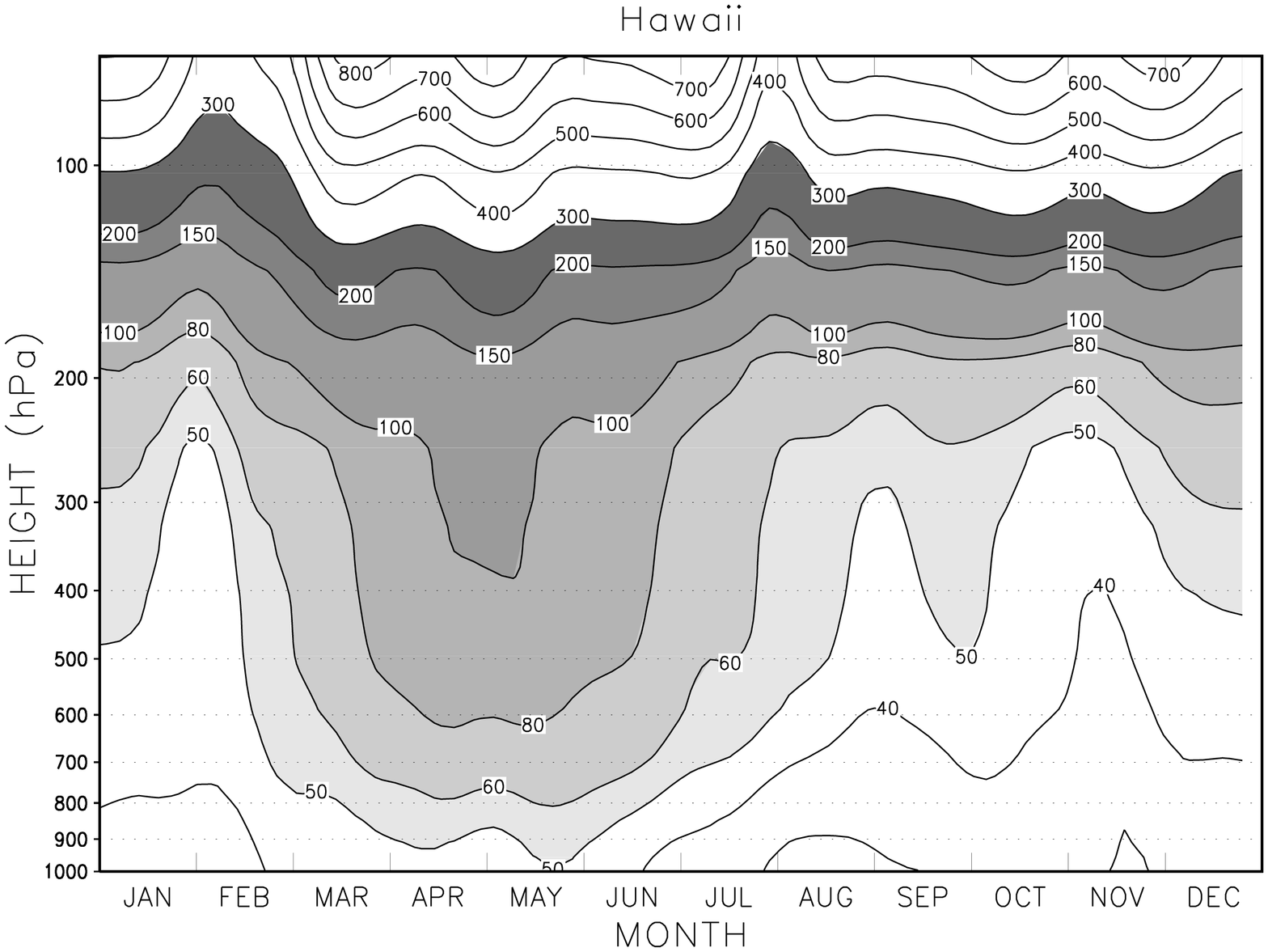}}
}
\vskip -0.0in
\centerline{
\leavevmode
(c)
\epsfxsize=1.5in
\epsfysize=3.0in
\rotatebox{-90.}{\epsfbox{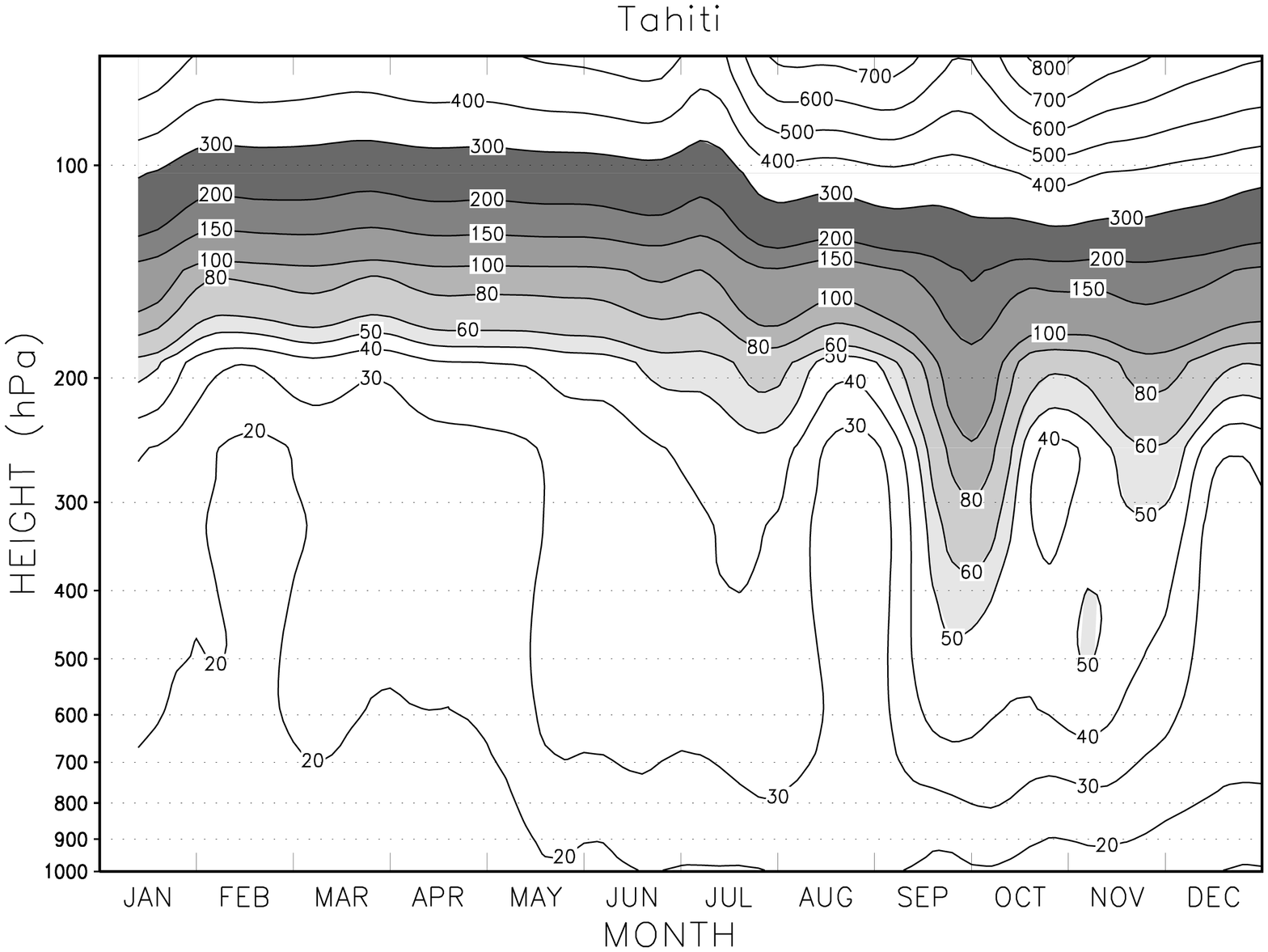}}
(d)
\epsfxsize=1.5in
\epsfysize=3.0in
\rotatebox{-90.}{\epsfbox{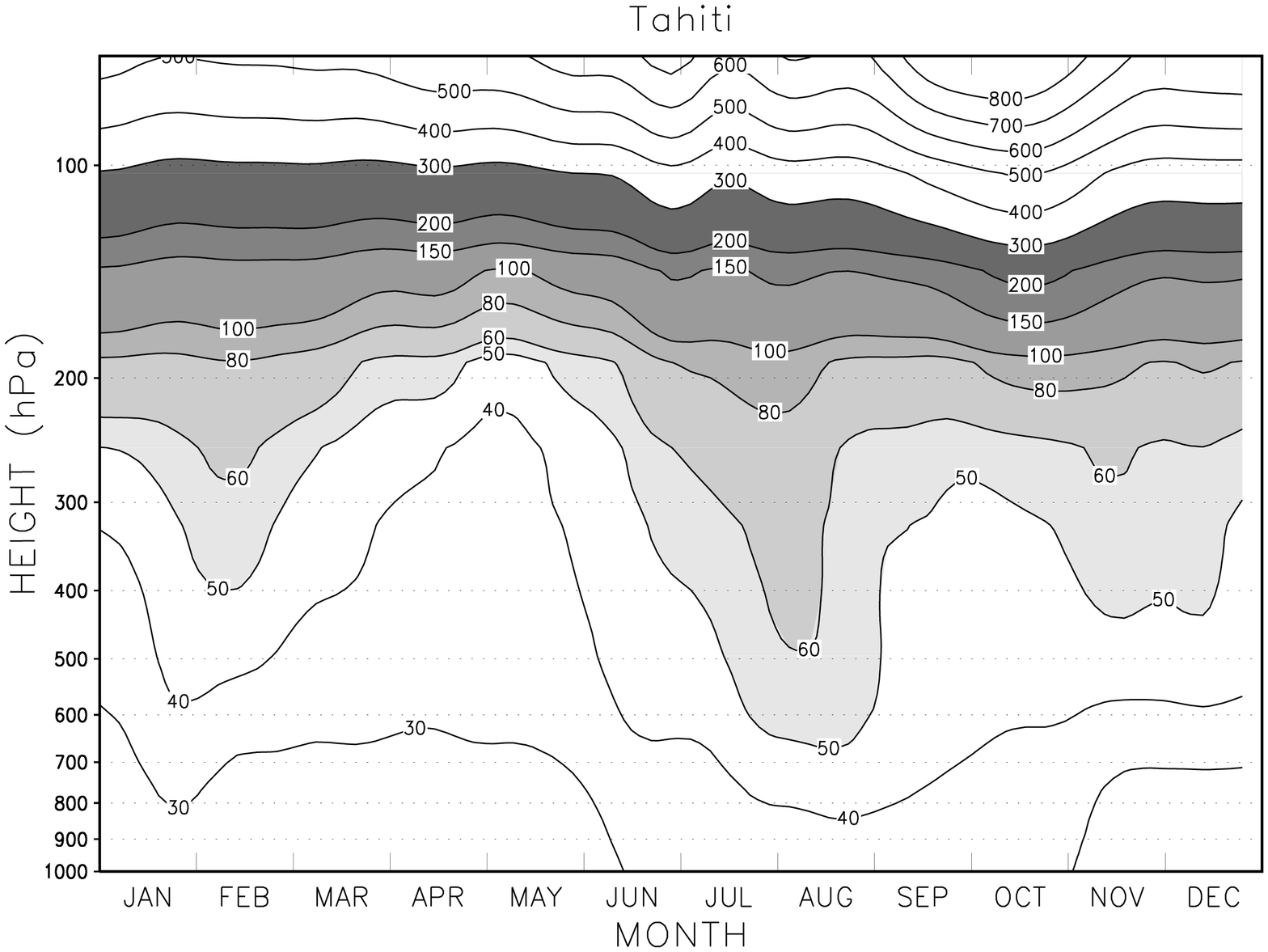}}
}
}
\caption{ \label{fig.ozonesonde.epa} Time-height cross sections of $O_3$ (ppbv) from measurements at  
Hawaii ($20^{\circ}N$, $155^{\circ}W$) (a) and Tahiti ($18^{\circ}S$, $149^{\circ}W$) (c). The model calculation for
these locations are shown in (b) and (d), respectively.}
\end{figure*}

The South Pacific is the region of the tropical troposphere most remote from human activity 
[{\it Schultz et al.}, 1999]. Figure~\ref{fig.ozonesonde.epa} (c) shows ozone vertical
profiles from ozonesondes at Tahiti. At this location, an identifiable SH spring
ozone maxima are seen as elevated ozone concentrations extend from upper troposphere downward
to the middle troposphere. The timing of this ozone maxima, from September to November, coinciding 
with the intensive SH biomass burning activities take place in South America, Africa, Southeast
Asia, and Oceania [{\it Schultz et al.}, 1999]. Elevated ozone ($\ge$ 40 ppbv) levels extend 
from the upper to the middle troposphere at Samoa in October. 

The measurements also exhibit
another period of elevated ozone ($\ge$ 40 ppbv) in the upper troposphere
in June-July. The model calculations show that elevated ozone extends from
the upper to the middle troposphere at this period, indicating the contribution of 
biomass burning emissions. However, the model does not show the elevated ozone as observed 
in September-October.
Modelled ozone vertical profiles at  
South Pacific sites (Java, Fiji, Samoa, and Tahiti) are persistently higher than
the observations in June-July, 
indicating that the model contains too much local biomass burning emissions, 
or too much biomass burning
is transported into this area from Africa or South America. On the other hand,
less biomass burning emissions has been produced or transported into South Pacific
during September-October.

For the NH site of Hawaii, both observed and modelled ozone profiles show
very distinctive seasonal spring maximum. Elevated ozone levels extend from
upper troposphere to near the surface. Seasonal minima appear during the summer months.
{\it Wang et al.} [1998] attributed this strong spring maximum to the long-range
transport of Asian pollution over the North Pacific. Analyses of anthropogenic
aerosols [{\it Perry et al.}, 1999] and CO [{\it Jaffe et al.}, 1997]
at this site also shows a seasonal maximum in spring with sources from
Asia continent [{\it Perry et al.}, 1999].

In summary, the comparison of time series ozone vertical distribution 
at Pacific basin shows an interhemispheric asymmetry in
ozone concentrations between NH and SH subtropics. While the NH spring maxima
are maintained by the continental outflow and long-range transport of continental
anthropogenic pollutants, the SH spring maxima are likely due to the biomass 
burning emissions which take place in Southeast Asia, Oceania, southern Africa,
and South America, and transport of $O_3$ precursors such as PAN and NOx from
soil and lightning [{\it Schultz et al.}, 1999].

\subsubsection{3.3.6. Northern Higher Latitudes}

\begin{figure*}[hp]
\vbox{
\vskip -0.0in
\centerline{
\leavevmode
(a)
\epsfxsize=1.5in
\epsfysize=3.0in
\rotatebox{-90.}{\epsfbox{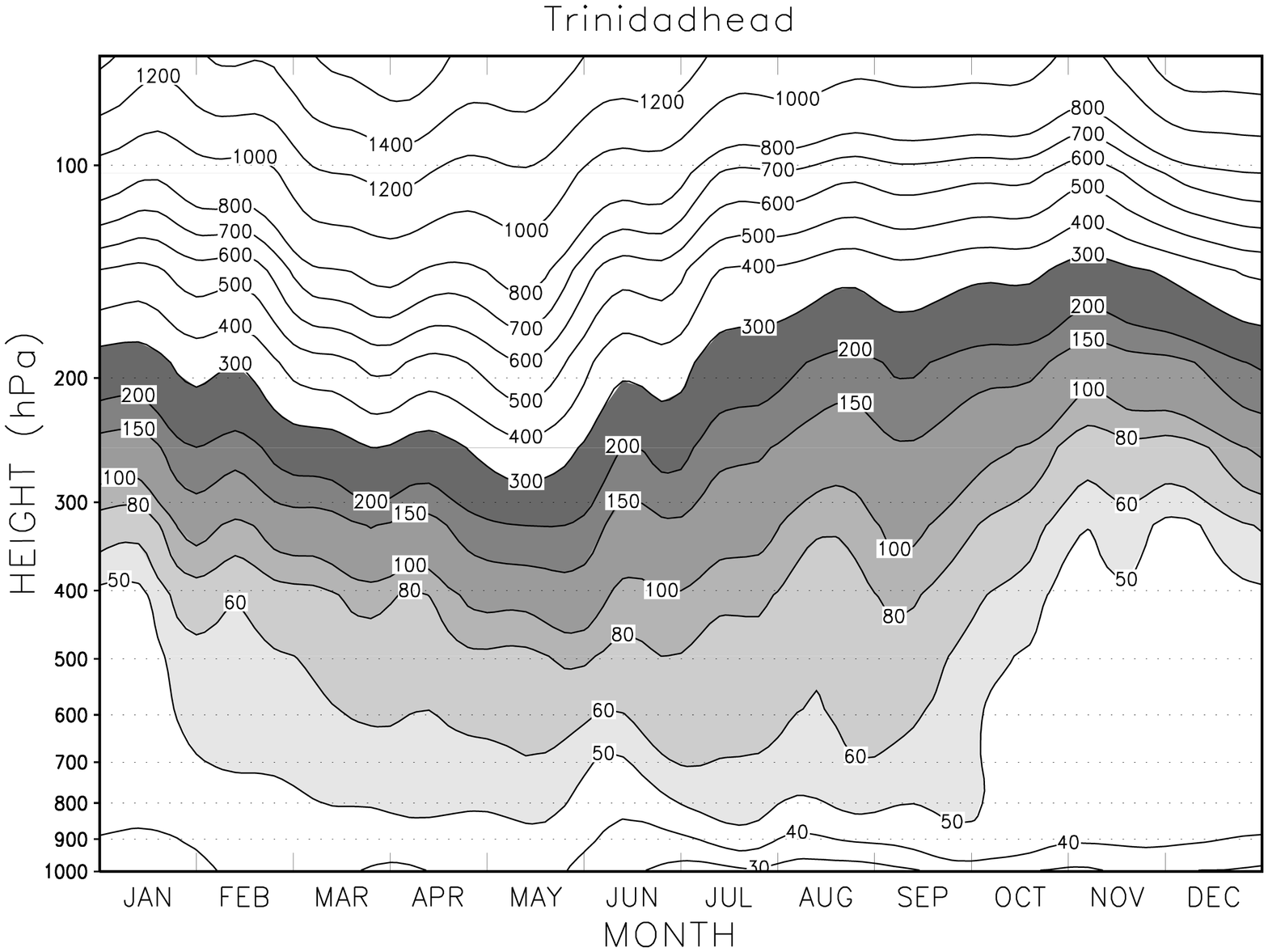}}
(b)
\epsfxsize=1.5in
\epsfysize=3.0in
\rotatebox{-90.}{\epsfbox{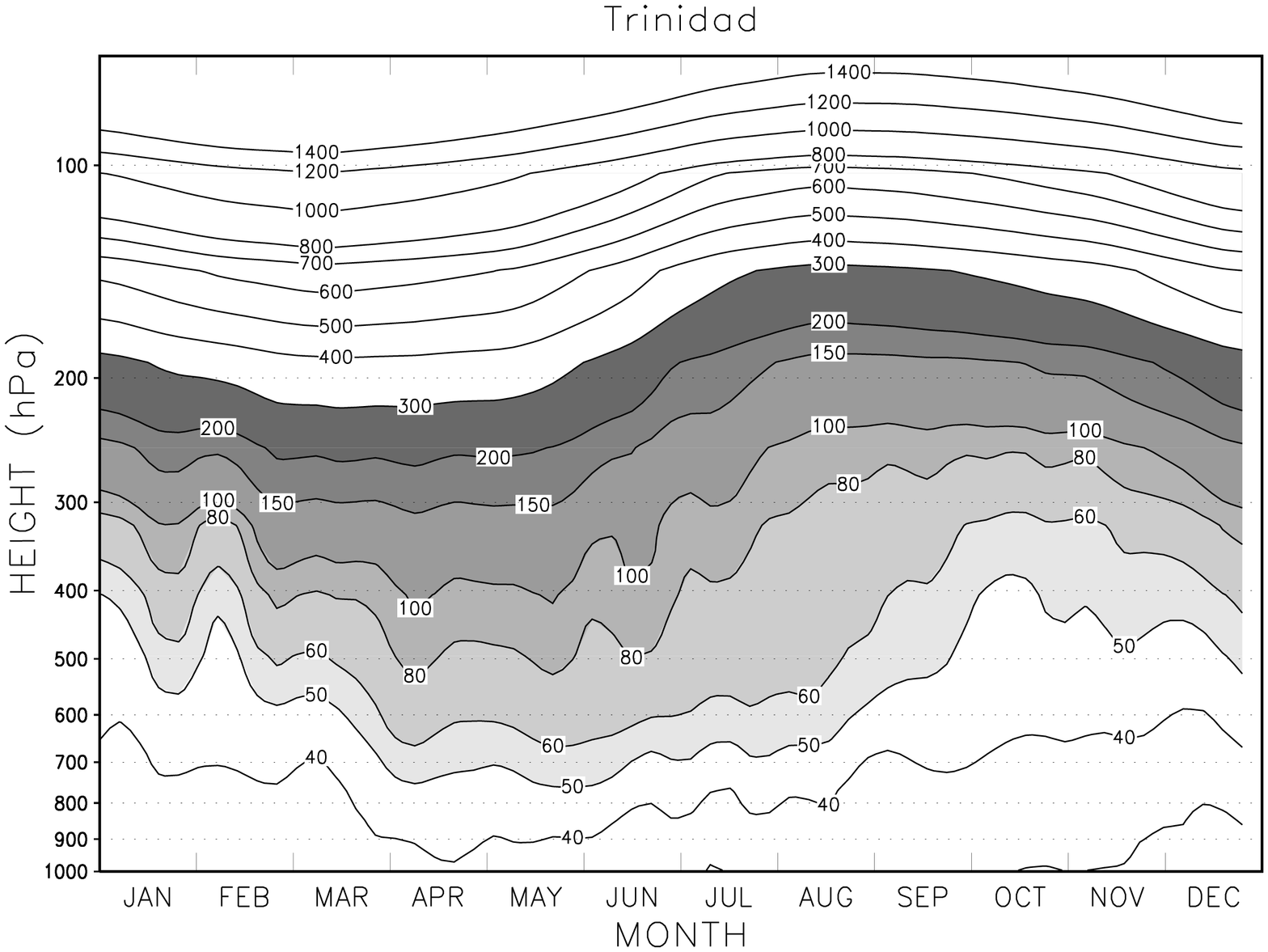}}
}
\vskip -0.0in
\centerline{
\leavevmode
(c)
\epsfxsize=1.5in
\epsfysize=3.0in
\rotatebox{-90.}{\epsfbox{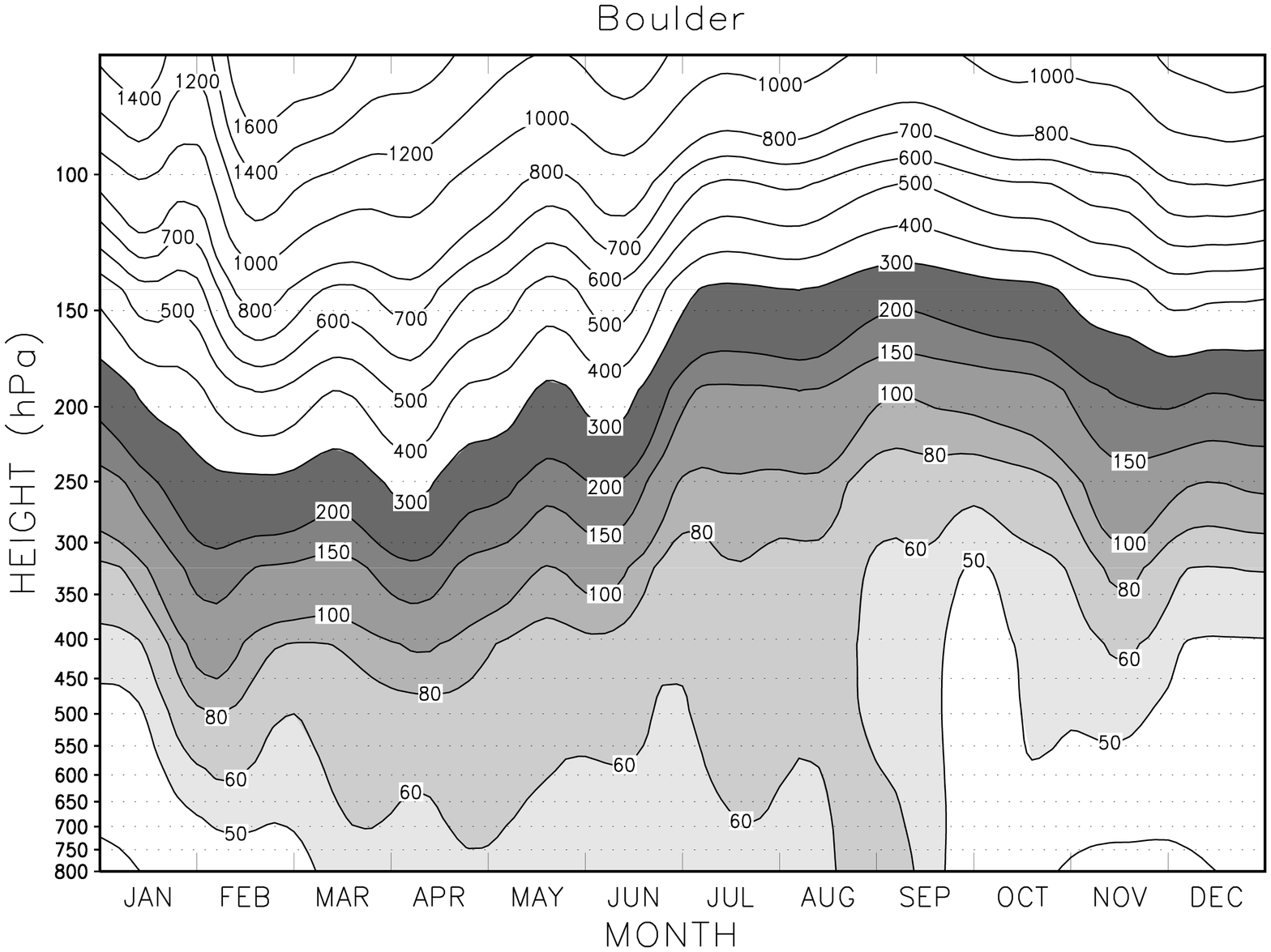}}
(d)
\epsfxsize=1.5in
\epsfysize=3.0in
\rotatebox{-90.}{\epsfbox{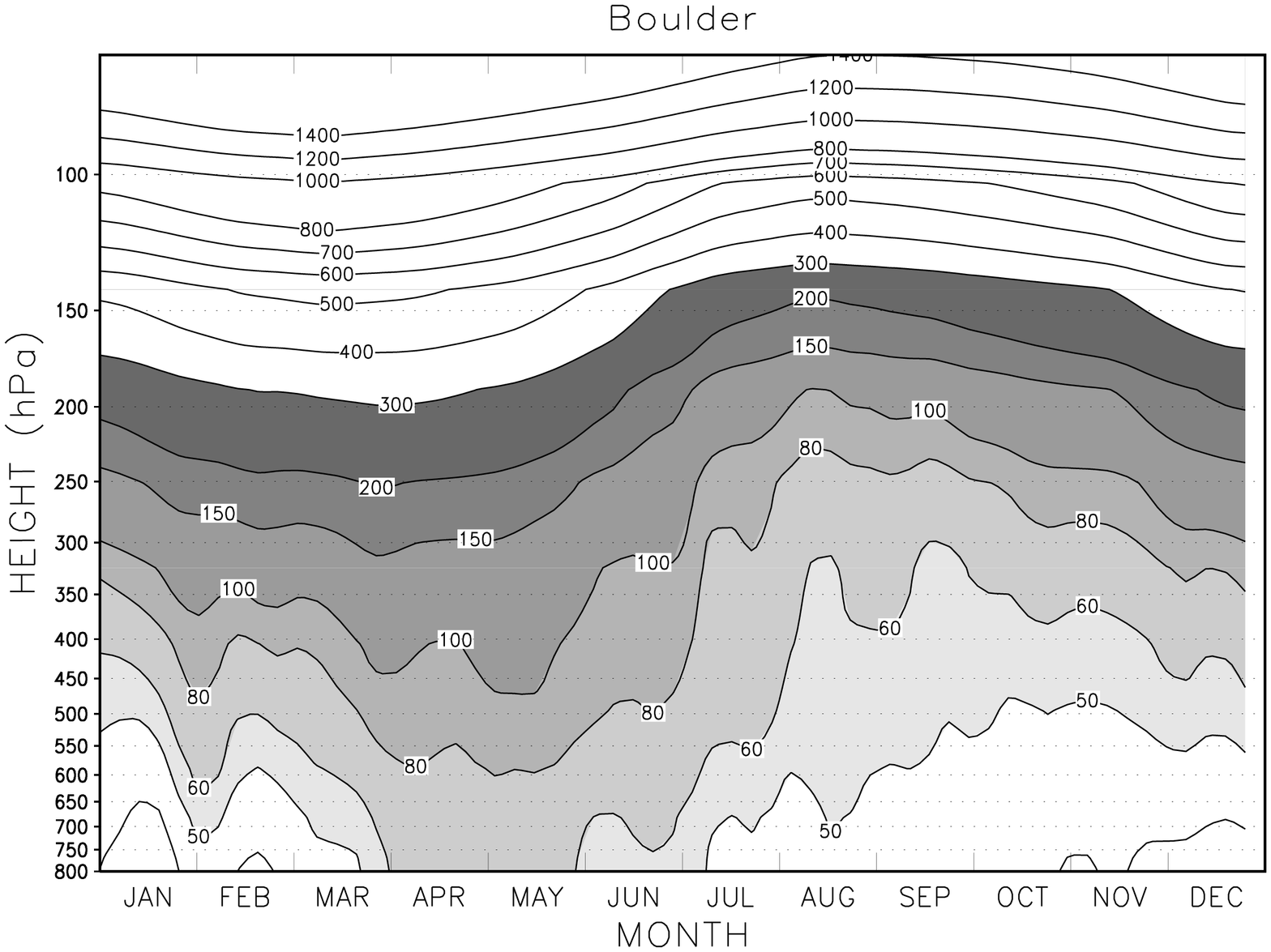}}
}
}
\caption{ \label{fig.ozonesonde.ohl} Time-height cross sections of $O_3$ (ppbv) from measurements at  
Trinidadhead ($41^{\circ}N$, $124^{\circ}W$) (a) and Boulder ($40^{\circ}N$, $105^{\circ}W$) (c). The model calculation for
these locations are shown in (b) and (d), respectively.}
\end{figure*}

Figure~\ref{fig.ozonesonde.ohl} extends the comparison of ozone vertical distribution from ozonesonde
measurements with model to higher latitude locations in the NH.
Both model and measurements at 
Trinidadhead and Boulder show highest ozone in spring in the lower stratosphere (200 hPa),
and in April to August in the middle troposphere and near the surface. Limited ozonesonde
measurements and model calculations at Fairbanks show similar timing for the occurrence of
spring ozone maximum compared with Trinidadhead and Boulder. 
Model calculations at the Azores show elevated ozone in the lower stratosphere in spring,
in the middle troposphere in spring to summer, and in spring in the lower troposphere.
These characteristics, high ozone extending downward from the tropopause region, 
are close to the available measurements [{\it Oltmans et al.}, 1996].

%
% 4. Summary
%
\section{4. Summary}

In this papers
we present modelling
results from a 3D CTM for the global troposphere. The modelled results
are examined at the surface and on a series of time-height cross sections
at several locations spread over the Atlantic, the Indian, and the Pacific.
Comparison of model with surface measurements at remote MBL stations
indicate a close agreement. The most striking feature of the hemispheric
spring ozone maximum in the MBL can be most easily identified from these 
NOAA CMDL surface ozone measurements, at the NH sites of Westman Island,
Bermuda, and Mauna Loa, and at the SH site of Samoa. 
Modelled ozone vertical distribution in the troposphere are compared with 
ozone profiles from NOAA CMDL and NASA SHADOZ ozonesonde measurements. 
For the Atlantic and the Indian sites, 
the model generally produces a hemispheric spring
ozone maximum close to those of the measurements.
The model also produces the spring ozone maximum
in the northeastern and tropical north Pacific close to those measurements,
and at sites in the NH high latitudes. 
The close agreement between model and the measurements indicate that the model
can reproduce the proposed processes responsible for producing the
spring ozone maximum in these regions of the MBL, 
lending confidence in the use of the model to investigate MBL ozone chemistry.
Overall, the model appears to perform better at sites where stratospheric
and biomass burning emissions is the major contributor. For example,
for sites at the Atlantic basin, western Indian (except Nairobi),
central north Pacific,
and the NH high latitudes. The model performance is degraded as the
distance between the site and the continental source increase, or as
the site located close to the equator (e.g., San Cristobal and Nairobi),
where the complex tropical circulations (e.g., Walker circulations and 
Hadley circulations) and deep cloud convections are more difficult for
a model to handle well. Other factors such as dry deposition, heterogeneous
chemistry, halogen chemistry, model resolution, and 
cross tropopause transport can all affect model results and need
further investigations.
In the following two papers (Parts 2 and 3) we investigate the impact of 
stratosphere-troposphere exchange 
and biomass burning emission on the
simulated ozone distribution, respectively.

\acknowledgments
The authors like to thank BADC, ECMWF, Central Weather Bureau (Taiwan),
S.J. Oltmans,
W.-S. Kau, G. Carver, and Brian Doty for their support on this work. 
We are very grateful to the NASA SHADOZ project for the
ozonesonde data archive ($http://hyperion.gsfc.nasa.gov/Data services/shadoz/Sites2.html$). 
This research was supported by the NSC grant NSC-89-2119-M-008-007. 
The Centre for Atmospheric Science is a joint initiative of the
Department of Chemistry and the Department of Applied Mathematics and
Theoretical Physics. This work forms part of the NERC U.K. Universities
Global Atmospheric Modeling Programme.

\balance

\end{document}